\documentclass[twocolumn]{aastex62}

\graphicspath{{./}{figures/}}

\newcommand{\ang}{\, \text{\AA}}

\accepted{for publication in ApJ, 2019 November 25}

\shorttitle{Ultra Massive Quiescent Galaxies at $z\gtrsim2$}
\shortauthors{Stockmann et al}

\begin{document}

\title{X-shooter Spectroscopy and \textit{HST} Imaging of 15 Ultra Massive Quiescent Galaxies at $z\gtrsim2$}

\author[0000-0001-5983-6273]{Mikkel Stockmann}
\affil{Cosmic Dawn Center (DAWN)}
\affil{Niels Bohr Institute, University of Copenhagen, Lyngbyvej 2, 2100 Copenhagen Ø, Denmark}
\affil{DARK, Niels Bohr Institute, University of Copenhagen, Lyngbyvej 2, DK-2100 Copenhagen, Denmark}
\email{mikkelstockmann@gmail.com}

\author[0000-0003-3631-7176]{Sune Toft}
\affil{Cosmic Dawn Center (DAWN)}
\affil{Niels Bohr Institute, University of Copenhagen, Lyngbyvej 2, 2100 Copenhagen Ø, Denmark}

\author[0000-0002-9656-1800]{Anna Gallazzi}
\affil{INAF-Osservatorio Astrofisico di Arcetri, Largo Enrico Fermi 5, I-50125 Firenze, Italy}

\author[0000-0003-1734-8356]{Stefano Zibetti}
\affil{INAF-Osservatorio Astrofisico di Arcetri, Largo Enrico Fermi 5, I-50125 Firenze, Italy}

\author[0000-0003-1949-7638]{Christopher J. Conselice}
\affil{Centre for Astronomy and Particle Theory, School of Physics $\&$ Astronomy, University of Nottingham, Nottingham, NG7 2RD, UK}
\affil{University of Nottingham, School of Physics $\&$ Astronomy, Nottingham, NG7 2RD, UK}

\author[0000-0001-8702-7019]{Berta Margalef-Bentabol}
\affil{LERMA, Observatoire de Paris, PSL Research University, CNRS, Sorbonne Universit\'es, UPMC Univ. Paris 06}

\author[0000-0002-9842-6354]{Johannes Zabl}
\affil{Univ Lyon, Univ Lyon1, Ens de Lyon, CNRS, Centre de Recherche Astrophysique de Lyon UMR5574, F-69230 Saint-Genis-Laval, France}

\author[0000-0003-3002-1446]{Inger J{\o}rgensen}
\affil{Gemini Observatory, 670 N.\ A`ohoku Pl., Hilo, HI 96720, USA}

\author[0000-0002-4872-2294]{Georgios E. Magdis}
\affil{Cosmic Dawn Center (DAWN)}
\affil{Niels Bohr Institute, University of Copenhagen, Lyngbyvej 2, 2100 Copenhagen Ø, Denmark}
\affil{Institute for Astronomy, Astrophysics, Space Applications and Remote Sensing, National Observatory of Athens, 15236, Athens, Greece}

\author[0000-0002-4085-9165]{Carlos G\'{o}mez-Guijarro}
\affil{Cosmic Dawn Center (DAWN)}
\affil{Niels Bohr Institute, University of Copenhagen, Lyngbyvej 2, 2100 Copenhagen Ø, Denmark}
\affil{DARK, Niels Bohr Institute, University of Copenhagen, Lyngbyvej 2, DK-2100 Copenhagen, Denmark}

\author[0000-0001-6477-4011]{Francesco M. Valentino}
\affil{Cosmic Dawn Center (DAWN)}
\affil{Niels Bohr Institute, University of Copenhagen, Lyngbyvej 2, 2100 Copenhagen Ø, Denmark}
\affil{DARK, Niels Bohr Institute, University of Copenhagen, Lyngbyvej 2, DK-2100 Copenhagen, Denmark}

\author[0000-0003-2680-005X]{Gabriel B. Brammer}
\affil{Cosmic Dawn Center (DAWN)}
\affil{Niels Bohr Institute, University of Copenhagen, Lyngbyvej 2, 2100 Copenhagen Ø, Denmark}

\author[0000-0002-8680-248X]{Daniel Ceverino}
\affil{Cosmic Dawn Center (DAWN)}
\affil{Niels Bohr Institute, University of Copenhagen, Lyngbyvej 2, 2100 Copenhagen Ø, Denmark}

\author[0000-0001-9197-7623]{Isabella Cortzen}
\affil{Cosmic Dawn Center (DAWN)}
\affil{Niels Bohr Institute, University of Copenhagen, Lyngbyvej 2, 2100 Copenhagen Ø, Denmark}
\affil{DARK, Niels Bohr Institute, University of Copenhagen, Lyngbyvej 2, DK-2100 Copenhagen, Denmark}

\author[0000-0002-2951-7519]{Iary Davidzon}
\affil{Cosmic Dawn Center (DAWN)}
\affil{Niels Bohr Institute, University of Copenhagen, Lyngbyvej 2, 2100 Copenhagen Ø, Denmark}
\affil{IPAC, California Institute of Technology, 1200 East California Boulevard, Pasadena, CA 91125, USA}

\author{Richardo Demarco}
\affil{Departamento de Astronom\'ia, Facultad de Ciencias F\'isicas y Matem\'aticas, Universidad de Concepci\'on, Concepci\'on, Chile}

\author[0000-0002-9382-9832]{Andreas Faisst}
\affil{IPAC, California Institute of Technology, 1200 East California Boulevard, Pasadena, CA 91125, USA}

\author[0000-0002-3301-3321]{Michaela Hirschmann}
\affil{DARK, Niels Bohr Institute, University of Copenhagen, Lyngbyvej 2, DK-2100 Copenhagen, Denmark}

\author[0000-0002-4912-9388]{Jens-Kristian Krogager}
\affil{Institut d'Astrophysique de Paris, UMR\,7095, CNRS-SU, 98bis bd Arago, 75014 Paris, France}

\author[0000-0003-3021-8564]{Claudia D. Lagos}
\affil{Cosmic Dawn Center (DAWN)}
\affil{International Centre for Radio Astronomy Research (ICRAR), M468, University of Western Australia, 35 Stirling Hwy, Crawley, WA 6009, Australia}
\affil{ARC Centre of Excellence for All Sky Astrophysics in 3 Dimensions (ASTRO 3D)}

\author[0000-0003-2475-124X]{Allison W. S. Man}
\affil{Dunlap Institute for Astronomy $\&$ Astrophysics, 50 St. George Street, Toronto, ON M5S 3H4, Canada}

\author[0000-0002-0833-8554]{Carl J. Mundy}
\affil{University of Nottingham, School of Physics $\&$ Astronomy, Nottingham, NG7 2RD, UK}

\author[0000-0001-8302-5198]{Yingjie Peng}
\affil{Kavli Institute for Astronomy and Astrophysics, Peking University, 5 Yiheyuan Road, Beijing 100871, China}

\author[0000-0001-9058-3892]{Jonatan Selsing}
\affil{Cosmic Dawn Center (DAWN)}
\affil{Niels Bohr Institute, University of Copenhagen, Lyngbyvej 2, 2100 Copenhagen Ø, Denmark}
\affil{DARK, Niels Bohr Institute, University of Copenhagen, Lyngbyvej 2, DK-2100 Copenhagen, Denmark}

\author[0000-0003-3780-6801]{Charles L. Steinhardt}
\affil{Cosmic Dawn Center (DAWN)}
\affil{Niels Bohr Institute, University of Copenhagen, Lyngbyvej 2, 2100 Copenhagen Ø, Denmark}
\affil{DARK, Niels Bohr Institute, University of Copenhagen, Lyngbyvej 2, DK-2100 Copenhagen, Denmark}

\author[0000-0001-7160-3632]{Kathrine E. Whitaker}
\affil{Department of Physics, University of Connecticut, Storrs, CT 06269, USA}
\affil{Department of Astronomy, University of Massachusetts, Amherst, MA 01003, USA}

 
\begin{abstract}
We present a detailed analysis of a large sample of spectroscopically confirmed ultra-massive quiescent galaxies (${\rm{log}}(M_{\ast}/M_{\odot})\sim11.5$) at $z\gtrsim2$. This sample comprises 15 galaxies selected in the COSMOS and UDS fields by their bright K-band magnitudes and followed up with VLT/X-shooter spectroscopy and \textit{HST}/WFC3 $H_{F160W}$ imaging. These observations allow us to unambiguously confirm their redshifts ascertain their quiescent nature and stellar ages, and to reliably assess their internal kinematics and effective radii. We find that these galaxies are compact, consistent with the high mass end of the mass-size relation for quiescent galaxies at $z=2$. Moreover, the distribution of the measured stellar velocity dispersions of the sample is consistent with the most massive local early-type galaxies from the MASSIVE Survey showing that evolution in these galaxies, is dominated by changes in size. The \textit{HST} images reveal, as surprisingly high, that $40\ \%$ of the sample have tidal features suggestive of mergers and companions in close proximity, including three galaxies experiencing ongoing major mergers. The absence of velocity dispersion evolution from $z=2$ to $0$, coupled with a doubling of the stellar mass, with a factor of four size increase and the observed disturbed stellar morphologies support dry minor mergers as the primary drivers of the evolution of the massive quiescent galaxies over the last 10 billion years.
\end{abstract}



\section{Introduction}

Local galaxies follow a bimodal distribution in color represented by blue star-forming spirals and red dormant elliptical galaxies. The most massive galaxies, primarily located in cluster environments, are the giant Elliptical galaxies with stellar population ages suggesting a formation more than 10 billion years ago \citep{Ma+14_massive,Greene+15}. 

A population of red massive galaxies are discovered to exist at $z\sim2$ \citep{Franx+03,Daddi+04} and subsequently confirmed to have quiescent stellar population \citep{Cimatti+04,Daddi+05,Labbe+05,Kriek+06a,Toft+07,Williams+09}. At this epoch the star formation rate density peaked \citep{Madau_Dickinson+14} alongside substantial nuclear activity (AGN) \citep{Hopkins+07}. At this time, half of the most massive (${{\rm{log}}}_{10}(M_{\ast}/M_{\odot})>11$) galaxies are already devoid of star formation (SF), and have old stellar ages suggesting that they quenched their star formation at even earlier times ($z>3$), when the Universe are only a few Gyr old \citep[e.g.][]{vanDokkum+06,Kriek+06b,Franx+08,vanDokkum+08,Toft+09,McCracken+10,Williams+10,Wuyts+11,Brammer+11,Whitaker+11,Kado_Fong+17,Morishita+18}. Nowadays quiescent galaxies are popularly defined by the \textit{UVJ} color-color relations \citep[see e.g.][]{Muzzin_COSMOS_Uvista}.

These massive quiescent galaxies are found to be remarkably compact with extremely high stellar densities when compared to local galaxies with similar stellar mass \citep{Papovich+05,Trujillo+06,Trujillo+07, Buitrago+08,vanDokkum+08,Cimatti+08,Bezanson+09,Conselice+11,Szomoru+12,vanderWel+14,Mowla+18}. A small number of elliptical galaxies this compact are found in the local Universe \citep{Trujillo+09,Taylor+10b,Shih_stockton+11,Ferre_Mateu+12}, but these are too young (ages $\sim2-4$ Gyr) to be the descendants of $z=2$ compact quiescent galaxies. This suggests that the vast majority of the $z=2$ population must undergo a substantial increase in size to evolve into local elliptical galaxies \citep{Bell+12}.

\cite{Bluck+12} found that the expected size evolution between $z=2.5$ and present day can be described primarily by minor mergers. However \cite{Newman+12,Man+16} found that minor mergers can account for the evolution at $z<1$ and that additional mechanisms of growth is required at higher redshift. The minor merger scenario is supported by the continuous size evolution found in compilation of spectroscopic \citep{Damjanov+11,Belli+14,Matharu+19} and photometric \citep{vanderWel+14,Faisst+17,Mowla+18} studies as well as the expected theoretical predictions of the galaxy properties during merger evolution \citep[e.g.][]{Khochfar_Silk+06,Naab+09,Lagos+18}.

\begin{figure*}[t!]
	\centering
	\includegraphics[width=18cm]{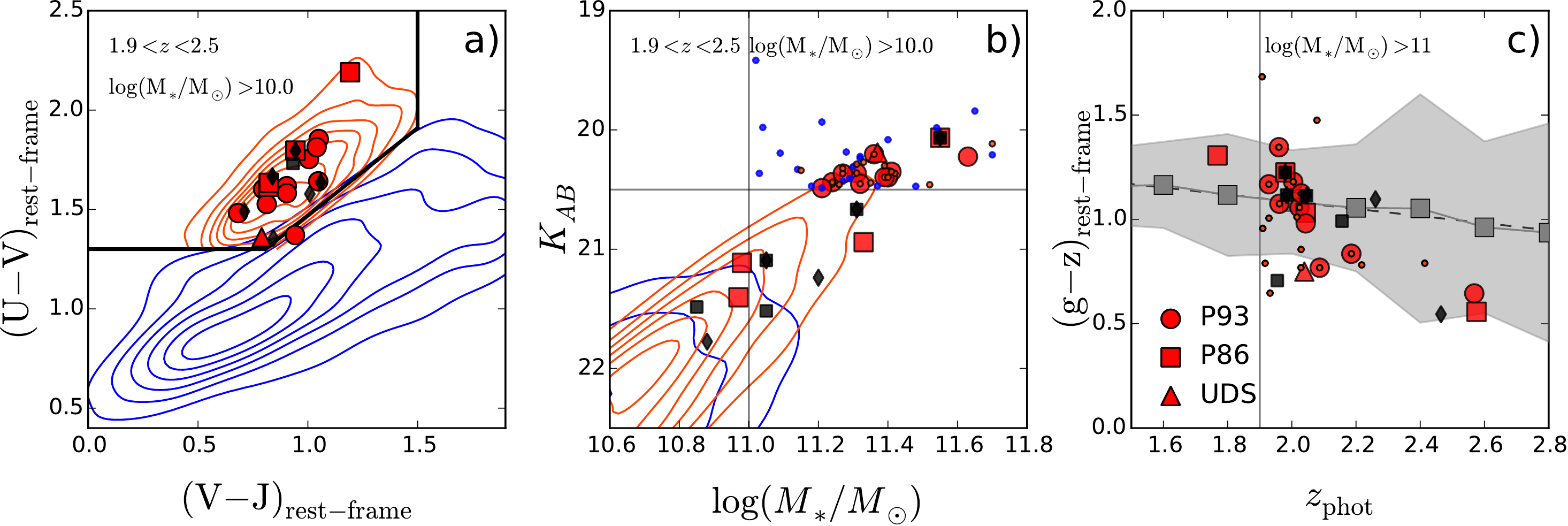}
    	\caption{Photometric properties of the galaxy sample (red 
        symbols - see legend on right) in the \textit{UVJ} (a), the $K_{AB}$-${{\rm{log}}}(M_{\ast}/M_{\odot})$ (b), and $z_{phot}$- rest-frame $\rm{(g-z)}$ planes from the \cite{Muzzin_COSMOS_Uvista} catalog. Note that for UDS19627 we use the \cite{Toft+12} K-band, stellar mass, $z_{phot}$ and rest-frame colors estimated from the observed photometry with EAZY \citep{Brammer+11}. The \textit{UVJ} quiescent (red) and star-forming (blue) galaxies are shown in contours in the range $1.9 < z_{phot} < 2.5$ and ${{\rm{log}}}(M_{\ast}/M_{\odot})>10$ \citep{Muzzin_COSMOS_Uvista}. The spectroscopically confirmed $z>2$ MQGs from COSMOS are shown with black symbols (square: \cite{Krogager+14}, diamond: \cite{Belli+17}). The small red/blue points in (b) are the galaxies that satisfy the criteria $K<20.5$ and ${{\rm{log}}}(M_{\ast}/M_{\odot})\rm{>11}$. The gray squares in panel (c) represent the running mean of the rest-frame $\rm{(g-z)}$ color of the massive, ${\rm{log}}_{10}(M_{\ast}/M_{\odot})>11$, \textit{UVJ}-selected quiescent galaxies with the $1\sigma$ standard deviation in gray.} \label{fig:selection}
\end{figure*}


To study the dynamics of massive quiescent galaxies at $z>2$, it is important to obtain both reliable kinematic and morphological measurements using deep spectroscopic observations and high resolution (adaptive optics or space-based) imaging \citep{Kriek+09,Toft+12,vandeSande2013,Belli+17}. Quiescent galaxies beyond $z>2$ are more disk-like with higher ellipticities than local ellipticals \citep{Toft+05,Toft+07,vanderWel+11,Wuyts+11}, which may cause heightened dispersion measurements from the contribution of unresolved rotation. In \cite{Toft+17} and \cite{Newman+18}, the first spatially resolved gravitationally lensed $z>2$ massive quiesecent galaxy are observed.\\

Massive quiescent galaxies at $z\sim2$ are rare \citep{Arcila-Osejo+19} and their quiescent nature implies faint rest-frame UV continua with no strong emission lines. Due to their rarity, large survey fields are essential to locate these galaxies. So far only a small samples of massive quiescent galaxies have been spectroscopically confirmed at $z>2$, in existing surveys like CANDELS+GOODS, and few of those have robust velocity dispersion measurements \citep{vandeSande2013,Belli+14,Kriek+16,Belli+17,Morishita+18}.


In this paper, the structural and dynamical properties of $15$ \textit{UVJ} massive quiescent galaxies (MQGs), ${\rm{log}}_{10}(M_{\ast}/M_{\odot})>11$, at $z>2$ are studied, doubling the spectroscopically confirmed and absorption-line detected sample at this epoch using the 2 sq. deg. COSMOS and UDS field. These MQGs are examined in detail through their evolution to local galaxies and how they likely formed in minor and major merger processes. In a follow-up paper, the Fundamental Plane relation and its evolution to $z=0$ is studied \citep{Djorgovski&Davis+87,Dressler+87}.\\
\linebreak{}
In Section \ref{sec:sample_selection}, we present the sample selection of the $z=2$ galaxies and a corresponding local reference sample. The X-shooter spectroscopic and \textit{HST} imaging data reduction, alongside the photometry used throughout the paper, are presented in Section \ref{sec:data}. In Section \ref{sec:analysis} we present the methods used to extract the X-shooter absorption-line kinematics, stellar populations and the \textit{HST} structural properties from the data, together with a multi-wavelength comparison of different star formation tracers. We address the issue of progenitor bias using our local reference sample in Section \ref{sec:minimal_progenitor_bias}. We present the stellar population, kinematic and structural results in Section \ref{sec:dispersion_size} and \ref{sec:mass_size}, and the dynamical properties in Section \ref{sec:dynamicalmass}. The results and the evolution of these galaxies to $z=0$ are discussed and summarised in Section \ref{sec:sum_discussion} and \ref{sec:summary}, respectively.\\
\linebreak{}
Throughout the manuscript, magnitudes are quoted in the AB system \citep{Oke_n_Gunn+83,Fukugita+96}, and the following cosmological parameters are used: $\Omega_{m}=0.3$, $\Omega_{\Lambda}=0.7$, with $H_{0}=70$ km/s/Mpc. All stellar masses are presented using the \cite{Chabrier+03} Initial Mass Function (IMF).


\begin{table*}[t!]
\centering{
    \caption{Summary of Sample}
    \label{tab:spec_info}
    \begin{tabular}{llllllllll}
\hline
Target ID & RA [degree] & Dec [degree] & $z_{\rm{phot}}$ & Exp. time & $K$ & $S/N_{H_{AB}}$ & ESO Program & $(U-V)$ & $(V-J)$ \\ \hline
UV-108899               & 150.17661   & 2.0608871   & $2.19$   & 5.0   & 20.35   & 5.69  & 093.B-0627(A) & 1.60 & 0.80  \\
UV-250513               & 149.82227   & 2.6531196   & $2.03$   & 5.0   & 20.37   & 4.12  & 093.B-0627(A) & 1.58 & 0.90  \\
CP-561356               & 150.20888   & 1.8502616   & $2.58$   & 5.6   & 20.94   & 2.16  & 086.B-0955(A) & 1.63 & 0.82  \\
UV-105842               & 150.26265   & 2.0177791   & $1.93$   & 4.0   & 20.20   & 4.28  & 093.B-0627(A) & 1.75 & 1.01  \\
UV-171687               & 149.88702   & 2.3506956   & $2.04$   & 5.0   & 20.49   & 3.08  & 093.B-0627(A) & 1.37 & 0.94  \\
UV-90676$^\mathrm{b}$   & 150.48750   & 2.2700379   & $2.57$   & 5.0   & 20.22   & 5.34  & 093.B-0627(A) & 1.53 & 0.81  \\
CP-1291751              & 149.86954   & 2.3167057   & $1.77$   & 7.2   & 21.40   & 1.80  & 086.B-0955(A) & 2.19 & 1.19  \\
UV-155853               & 149.55630   & 2.1672480   & $1.96$   & 5.0   & 20.36   & 4.65  & 093.B-0627(A) & 1.85 & 1.05  \\
UV-171060$^\mathrm{a}$  & 149.78951   & 2.3413286   & $2.02$   & 5.0   & 20.45   & 3.89  & 093.B-0627(A) & 1.62 & 0.90  \\
UV-230929               & 150.20842   & 2.7721019   & $2.09$   & 6.0   & 20.44   & 6.46  & 093.B-0627(A) & 1.48 & 0.68  \\
UV-239220               & 149.43275   & 2.5106428   & $2.00$   & 4.5   & 20.40   & 2.86  & 093.B-0627(A) & 1.64 & 1.05  \\
UV-773654               & 150.74574   & 2.0104926   & $1.96$   & 5.0   & 20.40   & 2.97  & 093.B-0627(A) & 1.81 & 1.04  \\
CP-1243752$^\mathrm{c}$ & 150.07394   & 2.2979755   & $1.98$   & 4.5   & 20.07   & 5.25  & 086.B-0955(A) & 1.80 & 0.94  \\
CP-540713               & 150.32512   & 1.8185385   & $2.04$   & 4.8   & 21.11   & 2.98  & 086.B-0955(A) & 1.61 & 0.82  \\
UDS19627$^\mathrm{d}$   & 34.57125    & -5.3607778   & $2.02$   & 5.0   & 20.19  & 4.40   & X-shooter GTO & 1.36 & 0.79 \\ \hline
\end{tabular}
}
\begin{flushleft}
\tablecomments{Target ID, right ascension (RA), declination (Dec), photometric redshift, X-shooter near-IR arm exposure time in hours, Total $K$ magnitude, median S/N ($9$ \AA{}/pixel bins) in H-band ($15000 < \lambda[\rm{\AA{}}] < 18000$), ESO program ID, rest-frame $(U-V)$ and $(V-J)$. The RA, Dec, photometric redshift, K-band, and UVJ colors are from \cite{Muzzin_COSMOS_Uvista} (except UDS19627$^\mathrm{d}$).}
\tablenotetext{a}{Previously published in \cite{Mowla+18}} 
\tablenotetext{b}{Previously published in \cite{Kado_Fong+17,Marsan+18,Mowla+18}} 
\tablenotetext{c}{Previously published in \cite{vandeSande2013,Krogager+14,Belli+14,Allen+15,Kriek+16,Belli+17,Mowla+18}} 
\tablenotetext{d}{Previously published in \cite{Toft+12} (all values in table taken from there)} 
\end{flushleft}
\end{table*}

\section{Sample selection} \label{sec:sample_selection}
The sample studied here consists of 15 MQGs from the COSMOS and UDS \citep{Williams+09} fields for spectroscopic follow-up and is selected based on the modeling of their optical to far-infrared broadband SEDs. Three samples, from three periods of observation, are presented below.
In the first program, galaxies were identified to be at $z_{\mathrm{phot}} > 1.6$ and with old ($>1$Gyr), quiescent stellar populations (specific star formation rates $\rm{log(sSFR/yr)}<-11$) in the updated version of the \cite{Ilbert+09} catalog of the COSMOS field described in \cite{Man+12}. The four $K$ band brightest ($K<21.5$) sources covered by parallel \textit{HST}/NICMOS observations were selected for follow-up to enable study of their morphology. These galaxies are referred to as the \textit{P86} sample, named after the period of VLT/X-shooter observations (P86, 2010-2011).\\

In a second program, 10 of the $K$ band brightest ($K<20.5$) galaxies in the COSMOS field with photometric redshifts\footnote{using redshift quality parameter with odds=1 from \cite{Brammer+08}} $z_{\mathrm{phot}} > 1.9$, specific starformation rates $\rm{log(sSFR/yr)} < -10$, and stellar masses ${\rm{log}}_{10}(M_{\ast}/M_{\odot})>11$ from the \cite{Muzzin_COSMOS_Uvista} catalog were selected for follow-up. 
Based on visual inspection, the sources with nearby bright objects in the $K$ band images are excluded to avoid photometric contamination. Objects with \textit{Spitzer}/MIPS $24\ \mathrm{\mu{m}}$ detections are also excluded to avoid either dusty star-forming galaxies or AGN  \citep{Le_Floch+09}. Their SEDs were visually inspected and galaxies with noisy photometry or bad fits were excluded. This pool of galaxies are dubbed the \textit{P93} sample, observed 3 years after P86.\\

Finally, in the analysis presented here, the massive quiescent galaxy UDS19627, from \cite{Toft+12}, is included. This object are selected as part of early VLT/X-shooter GTO observations to be quiescent ($\rm{log(sSFR/yr)} < -10$), at a high redshift ($z_{phot}=2.02^{+0.07}_{-0.08}$) and a bright source ($K=20.19$) in the UKIRT Ultra Deep Survey \citep{Williams+09}. New \textit{HST}/WFC3 $H_{F160W}$ imaging of this galaxy is presented, allowing us to measure resolved morphology. UDS19627 is minimally gravitationally lensed, but \cite{Toft+12} showed that, after taking this effect into account, the systematic change in magnification factor of $10-20\ \%$ correspond to a $0.07$ and $0.03$ dex resulting lower stellar and dynamical mass.\\

Our full sample is compiled from the three presented subgroups selected with variations in criteria on stellar mass, sSFR, and K-band brightness. In Figure \ref{fig:selection}a, we show that despite the variation in selection criteria, this sample populates the quiescent galaxy region of the \textit{UVJ} rest-frame color-color diagram \citep{Muzzin13b}. For the sake of homogeneity the full sample (except for UDS19627) is shown using the \cite{Muzzin_COSMOS_Uvista} catalog. Our galaxies are consistent with the UVJ selection for massive (${\rm{log}}(M_{\ast}/M_{\odot})>10$) quenched objects at $1.9 < z < 2.5$.

Figure \ref{fig:selection}b shows the position of our sample in the $K$-band magnitude - stellar mass plane. The $K<20.5$ and ${\rm{log}}(M_{\ast}/M_{\odot})>11$ selection of the P93 sample results in significantly larger stellar masses than the average for the P86 sample (selected as massive quiescent galaxies with NICMOS coverage) with only 1 galaxy from the latter fully satisfying the criteria of P93 \citep[previously presented in, among others,][]{vandeSande2013,Kriek+16,Belli+18}. The power of adding a minimum $K$-band threshold to the stellar mass criterion to select the most extreme massive quiescent galaxies is evident when comparing our sample with previous studies \citep{vandeSande2013,Krogager+14,Belli+17}, identifying on average massive quiescent galaxies with lower stellar masses. Our sample represents $60$\% of the total number of \textit{UVJ}-MQGs ($29$\% of all galaxies) at $1.9<z<2.5$, ${\rm{log}}(M_{\ast}/M_{\odot})>11$ and $K<20.5$ from \cite{Muzzin_COSMOS_Uvista} (upper right corner of Figure \ref{fig:selection}b). We confirm that our selection of \textit{UVJ} quiescent galaxies can be considered representative of the massive and K-band brightest galaxies at $1.9<z<2.5$. This is done by using a modified version of the Anderson-Darling test\footnote{\url{https://docs.scipy.org/doc/scipy/reference/generated/scipy.stats.anderson_ksamp.html}} to compare our stellar mass and K-band selection with the photometric samples respectively.

One concern addressed by \cite{vandeSande2014_FP} is that the selection of the K-band brightest galaxies introduces a bias towards the bluest galaxies in the rest-frame color $(g-z)_{rf}$. To address this issue the rest-frame colors $(g-z)_{rf}$, as a function of redshift between our sample and the \textit{UVJ} selected massive (${\rm{log}}(M_{\ast}/M_{\odot})>11$) quiescent galaxies from \cite{Muzzin_COSMOS_Uvista}, are compared in Figure \ref{fig:selection}c. 
Contrary to the sample of  \cite{vandeSande2014_FP}, 13/15 of our galaxies has $(g-z)_{rf}$ colors consistent within the standard deviation of the average massive quiescent galaxies at a matching epoch. The Anderson-Darling test for k-samples confirms that the $(g-z)_{rf}$ colors for our MQGs are representative of the (${\rm{log}}(M_{\ast}/M_{\odot})>11$) \textit{UVJ} massive quiescent galaxies at $1.9<z<2.5$. This suggests that, our K-band selected sample is on average not biased towards galaxies with bluer colors. However, the highest redshift sources have systematic lower $(g-z)_{rf}$ colors and could be subjected to this selection bias.

In summary, our sample is selected to be the most massive K-band bright \textit{UVJ} quiescent galaxies at $z>2$. The selection is not subjected to a bias in $(g-z)_{rf}$ and can be considered a $60\ \%$ stellar mass and K-band complete sample of the quiescent galaxies at $z>2$.

\subsection{A suitable reference sample of local galaxies} \label{sec:massive_n}

The MASSIVE Survey samples the most massive $K$-band selected early-type galaxies within the local 108 Mpc northern hemisphere \citep{Ma+14_massive}. These galaxies have central stellar ages suggesting a formation epoch at $z>2$ \citep{Greene+15}. Given the similar selection for our MQGs at $z>2$, stellar masses and inferred formation epoch, this sample is adopted as the local reference sample. This sample is further motivated in Section \ref{sec:minimal_progenitor_bias}.\\

The extinction-corrected absolute $K$-band magnitudes listed in Table 3 from \cite{Ma+14_massive} are converted into stellar masses using Equation (1) in \cite{vandeSande+19}. The NASA-Sloan Atlas semi-major axis optical effective radii, also listed in Table 3 from \cite{Ma+14_massive}, are used. These were derived from two-dimensional S\'ersic \citep{Sersic} fits with S\'ersic parameters varying between $n=2-6$. For the galaxies where this is not available, the infrared 2MASS measurements were used to convert these to semi-major axis optical effective radii using Equation (4) in \cite{Ma+14_massive}. These sizes were derived from single S\'ersic and de Vaucouleurs profile fits ($n=4$). The effective velocity dispersion measurements used are reported in \cite{Veale+18}. They were estimated using the MILES stellar library \citep{Falcon-Barroso+11} together with pPXF \citep{Cappellari_ppxf}. Finally, the average luminosity-weighted stellar velocity dispersion within the effective radius is adopted.


\section{Data} \label{sec:data}
Here, we describe the spectroscopic observations with the VLT/X-shooter spectrograph \citep{D'odorico+06,Vernet_2011_Xshooter} and the \textit{HST}/WFC3 follow-up of our MQGs. These spectroscopic and photometric campaigns spanned an interval of more than $10$ years, spread over several programs that are summarized in Table \ref{tab:spec_info}. Finally, the ancillary data used in the analysis are presented.\\

\begin{figure*}[!htbp]
  \includegraphics[width=18cm]{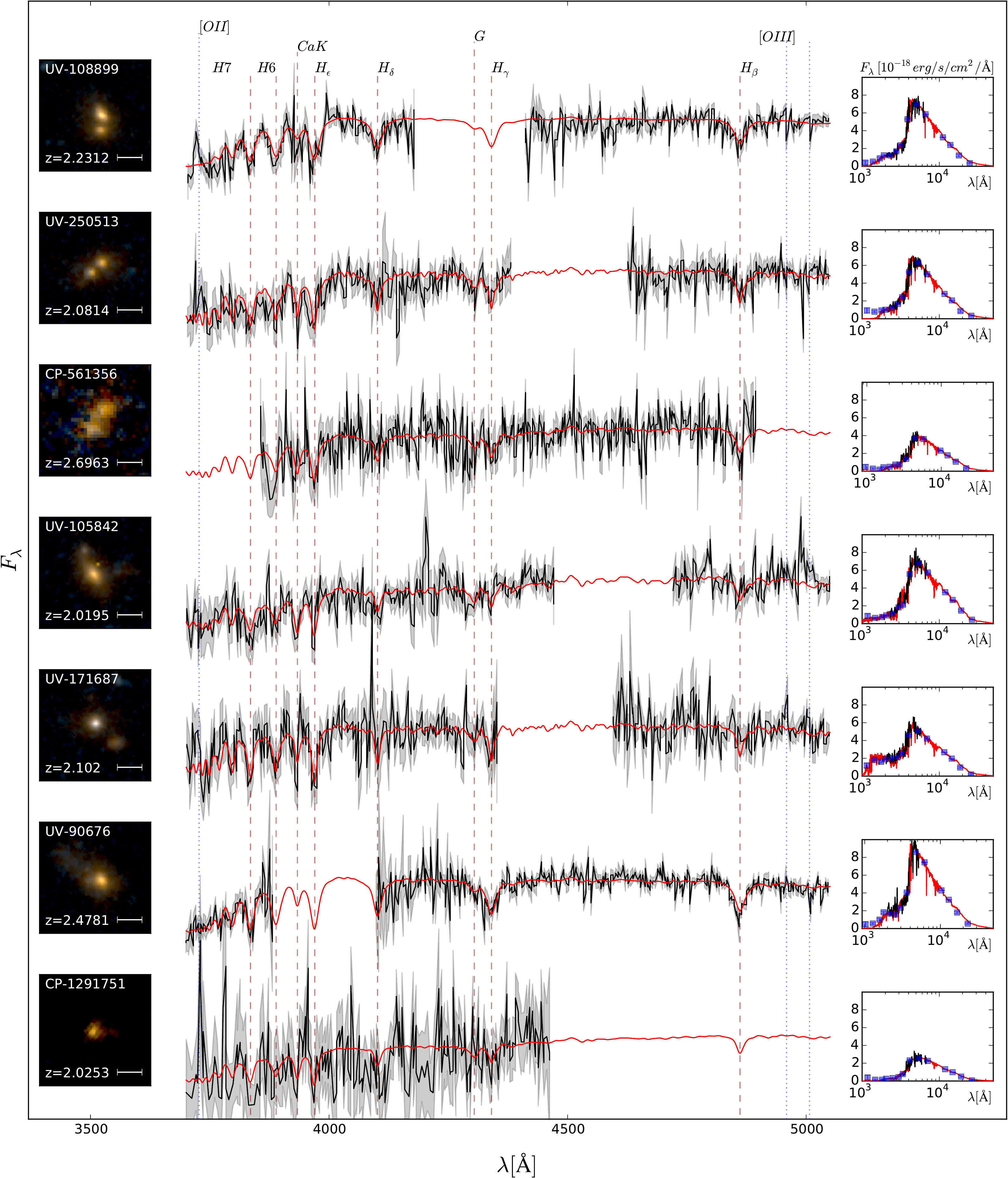}
    \caption{VLT/X-shooter spectra, of our sample, covering the rest-frame wavelength range $3700<\lambda{}<5050$ \AA{} with corresponding HST Red-Blue images, in the left column. To the right, the full SED displayed by multi-wavelength photometry (blue squares) and in center the rest-frame optical X-shooter spectra (black line) and the best-fit stellar population model (red line, Section \ref{sec:stelpop}). Spectra are shown with an optimal adaptive binning and $1\sigma$ rms noise in gray shading. The two-color $4.5"\times4.5"$ North-East orientated RB images, with galaxy ID and  absorption-line determined spectroscopic redshifts (determined in Section \ref{sec:veldisp}), are made from \textit{HST}/ACS $I_{F814W}$ and WFC3 $H_{F160W}$. A $1"$ white bar is shown ($\sim8.5$ kpc at $z=2$). The G, Ca K, and Balmer absorption features are indicated with dark red dashed lines while [OII]3727\AA\ and [OIII]4959, 5007\AA\ is indicated with blue dotted lines. The figure shows bright red sources with Balmer absorption lines, no significant optical emission lines, strong $4000$ \AA{} break, and low rest-frame UV light all indicative of quiescent stellar populations.}
    \label{fig:spec_presentation}
\end{figure*} 

\begin{figure*}[!htbp]
  \includegraphics[width=18cm]{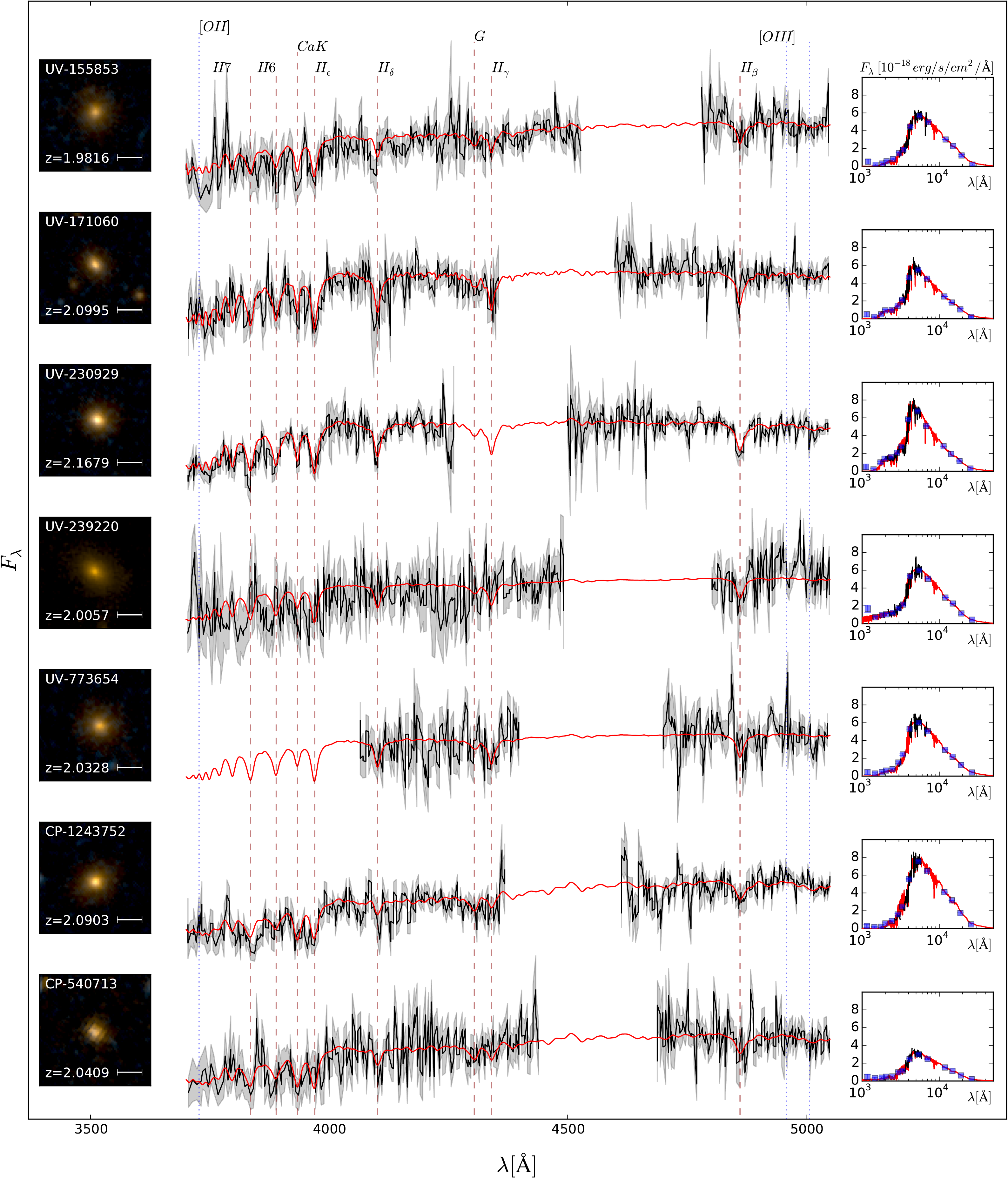}
  \centering{Figure \ref{fig:spec_presentation} - continued}
\end{figure*}

\subsection{VLT/X-shooter spectroscopy}
X-shooter is a single object Echelle spectrograph mounted on the VLT and covers $3,000 - 25,000 \ang$ with three arms: UVB ($2936-5930$ \AA{}), VIS ($5,253-10,489$ \AA{}), and NIR ($9,827-24,807$ \AA{}). We are granted 35 and 57 service mode hours in P86 and P93, respectively (PI: Toft). The latter carried over and finished in period 96. The observations are completed using default nodding mode to ensure a robust sky subtraction of the NIR band, probing the rest-frame optical part of the spectra for the $z\sim2$ quiescent galaxies. The majority of the P86/P93 observations ($89/96\%$) are completed with an average air-mass corrected DIMM seeing of $0\farcs8$ in the NIR arm. The telluric standard stars are observed close to the science observations, both in airmass and time to mimic the conditions of the sky and optimize the atmospheric absorption correction. The P86/P93 observations for the NIR (VIS) frames are executed with 480s/900s (314s/863s) exposures, $0.9"\times11"$ slit configuration and -- for the P93 sample only -- including the $K$-band blocking filter. We aligned the slit along the galaxy's major axis in the UltraVISTA K-band images avoiding bright nearby sources.\\

The data are reduced using a wrapper of the ESO X-shooter pipeline \citep{Modigliani_2010,XSHpipeline_Sparre}, along with customized modifications \citep{Zabl_himiko_2015}. Beyond the standard pipeline processing steps for the NIR arm in nodding mode, we account for the spatial variations of the background level outside of the orders in each raw science frame by removing the median level obtained from the illuminated areas from each row of pixels in the detector. The 2-D VIS and NIR individual science frames are corrected for telluric absorption with a customized and publicly available wrapper\footnote{\url{https://github.com/jselsing/QuasarComposite/blob/master/py/telluric.py}} \citep{Selsing_2016_telluric} of the Penalized Pixel-Fitting algorithm \citep[pPXF]{Cappellari_ppxf}, based on the PHOENIX stellar atmosphere library \citep{PHOENIX_stellar_lib}. A response function is constructed modeling the atmosphere during the science exposures and each individual observation block (OB) are corrected.

Finally, individual OBs are combined into an optimally weighted 2-D spectrum removing flux outliers using a $3$ and $5\sigma$ median clipping for the VIS and NIR, respectively. Bad pixels automatically flagged during the reduction are also excluded. Furthermore, off-trace emission is flagged and excluded in the construction of the OBs from UV-105842, UV-171687, and UV-155853 to minimize the contamination from surrounding sources. The 1-D spectrum is optimally extracted \citep{Horne_OptExt}. Flux corrections are made anchoring the synthetic photometry to the total magnitudes from the latest COSMOS15 catalog \citep{Laigle+16} (Section \ref{sec:data_photometry}), accounting for PSF matching in different bands and for the Galactic extinction. The $H$-band and $I$-band magnitudes are used to compute independent aperture correction factors for the NIR and VIS spectra, respectively.\\

\subsection{HST/WFC3 $H_{F160W}$ imaging}
11 orbits of \textit{HST}/WFC3 with HST-GO-14721 (PI: Conselice) are allocated to observe the rest-frame optical images, $H_{F160W}$, for UDS19627 and the 10 galaxies in the \textit{P93} sample. The \textit{P86} sample are covered by the following programs: CP-1243752 (HST-GO-12440, PI: Faber) and CP-561356  (HST-HLA-14114, PI: van Dokkum) with WFC3; CP-1291751 and CP-540713 with HST/NICMOS (HST-HLA-9999, PI: Scoville).\\

The WFC3/$H_{F160W}$ data is reduced using the ``Grism redshift and line'' analysis software, Grizli\footnote{\url{https://github.com/gbrammer/grizli/}}, which is an end-to-end processing code for WFC3/IR data using ASTRODRIZZLE\footnote{A Python implementation of Multidrizzle: \url{https://drizzlepac.readthedocs.io/en/latest/astrodrizzle.html}}. The starting point is the standard calibrated images downloaded from the MAST archive (FLT extension images). The calibrated images are $1014\times 1014$ pixels with $0''.13/\rm{pixel}$. For each visit, there are four dithered exposures that are combined using Grizli. The resulting products for each visit are aligned, background subtracted and drizzled images with $0''.06/\rm{pixel}$. The NICMOS data for CP-1291751 and CP-540713 are reduced in a similar manner with ASTRODRIZZLE.

\subsection{Ancillary data: Multi-wavelength photometry and \textit{HST} $I_{F814W}$ images} \label{sec:data_photometry}
We make ample use of the 14 broadband COSMOS photometry from the \cite{Laigle+16} catalog, covering the full UV-to-NIR wavelength range to model our stellar populations in Section \ref{sec:stelpop}. The total magnitudes are adopted using the method described in Appendix A.2 by the same authors. Complementary to the UV-to-NIR photometry, we check the available deep X-ray \textit{Chandra} imaging \citep{Marchesi+16_Chandra_cat_optIR} and the ``super-deblended'' far-infrared (FIR) catalog \citep{Jin+18}, superseding the previous 24 $\mu$m catalog \citep{Le_Floch+09} used in the selection of \textit{P93}. This new implementation adopts active priors from the \textit{Spitzer}/MIPS 24 $\mu$m and radio observations to deblend the low resolution imaging from \textit{Herschel}/PACS and SPIRE, SCUBA2, AzTEC, and MAMBO. The sources are cross-check with the GALEX far-UV and near-UV data from \cite{Zamojski+07} and \cite{Capak+07}. This search for UV or X-ray counterparts results in no detections for any of our galaxies. On the other hand, we do find hints of mid-infrared (MIR) and radio emission from part of the sample, as detailed in Section \ref{sec:mips24_sfr} and discussed in Section \ref{sec:disc_agn}. UDS19627 has similar UV-to-NIR multi-wavelength coverage. For an in-depth discussion of the available photometric data for this object, see \cite{Toft+12}.

$13/15$ galaxies have \textit{HST} $I_{F814W}$ imaging that are part of the COSMOS public released data \citep{Scoville+07,Koekemoer+07}. It covers $\sim2$ sq degrees of the sky with the Advanced Camera for Surveys (ACS) in the I-band and comprises 81 tiles. Each tile is observed in 4 dithered exposures that are combined to produce a pixel scale of 0\farcs03/pixel and a Point Spread Function (PSF) of 0\farcs095 at full width at half maximum (FWHM). COSMOS images reach a point source limiting depth of $AB(F814W) = 27.2\ (5\sigma)$.


\section{Analysis} \label{sec:analysis}

We present in this section the analysis of our X-shooter spectra and our \textit{HST}/WFC3 $H_{F160W}$ images. The spectroscopic redshift, the velocity dispersion and stellar population of our galaxies are measured by modeling the absorption features in the stellar continuum together with the broadband photometry. As we find no significant emission line detections in the spectra, we derive optical SFR upper limits (Section \ref{sec:SFR_quiescence}) which we compare with the estimates from the MIR photometry. The majority of the spatially offset sources caught in the spectra are foreground and background galaxies. Finally, the \textit{HST} images probing the rest-frame optical structure are modeled to obtain their morphological parameters. The major merger candidates (UV-108899, UV-250513, and CP-561356 - see Figure \ref{fig:spec_presentation}) are confirmed to be within redshift proximity such that their stellar masses reliably can be flux corrected.\\

The \textit{HST} Red-Blue (RB) color images, rest-frame optical X-shooter spectra with \citep{Laigle+16} photometry and our best fitting stellar population model is shown in Figure \ref{fig:spec_presentation}. For UDS19627, the HST/WFC3 $H_{F160W}$ image is presented in Section \ref{sec:galfit_modeling} and its spectrum is shown in \cite{Toft+12}.

\begin{figure}
  \centering
  \includegraphics[width=8.5cm]{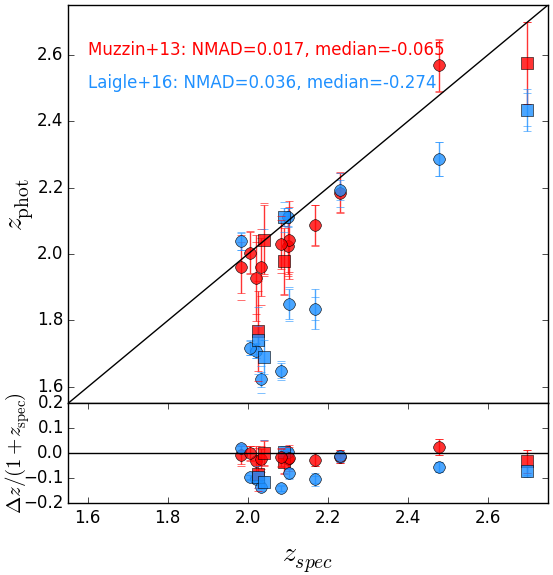}
     \caption{Comparison of spectroscopic and photometric redshifts for our sample of massive quiescent galaxies using the \cite{Muzzin_COSMOS_Uvista} (red) and \cite{Laigle+16} (blue) catalogs. The \cite{Muzzin_COSMOS_Uvista} catalog provides better photometric redshift estimates for massive quiescent galaxies at $z>2$ compared to \cite{Laigle+16}. Note that UDS19627 is not in the same area of the sky covered by the catalogs compare here.}
   \label{fig:redshift}
\end{figure}

\subsection{Spectroscopic redshifts and stellar velocity dispersion} \label{sec:veldisp}

All spectra of targeted sources (\textit{P86} and \textit{P93}) show prominent hydrogen absorption features, which are typical of evolved stellar populations (see Figure \ref{fig:spec_presentation}). The stellar absorption features are modeled using pPXF, and both the line of sight velocity centroid (i.e., the spectroscopic redshift) and the line of sight stellar velocity dispersion (LOSVD, hereafter ``velocity dispersion'') are measured.

The initial redshift and velocity dispersion guess is obtained from running pPXF with the \cite[]{BC+03} stellar population library (hereafter BC03). The stellar population analysis is performed with complex star formation histories (SFHs) fitting of the spectra and SED (see Section \ref{sec:stelpop}) adopting this initial estimate. The resulting best fit model is confirmed to be stable against perturbations of $\Delta{}\sigma=\pm{}100$ km/s. The velocity dispersion measurement is refined, by rerunning pPXF with a non-velocity broadened best-fit stellar population model.

The spectra and best-fit model are convolved to the same resolution ($FWHM=3.2$ \AA) and rebinned to a constant velocity scale without additional interpolation. Low order additive (a=2) and multiplicative (m=2) correction polynomials are fit over the rest-frame range $3750-5950$ \AA. The JH band gap and the regions, where emission lines might be expected\footnote{Excluded emission lines (wavelengths in \AA{}): [OII] (3726.03, 3728.82), [OIII] (4958.92, 5006.84), [OI] (6300.30), [NII] (6548.03, 6583.41), H$\alpha$ (6563), and [SII] (6716.47, 6730.85)}, are excluded while also masking out bad pixels.

The associated systematic and statistical errors are quantified by varying the wavelength range, correction polynomials, and stellar libraries (see details in Appendix \ref{app:dispersion_test}), similar to the method used in \cite{Toft+17}. In all cases (\textit{P86} and \textit{P93}), we determine secure redshifts and for $10/14$ galaxies we estimate robust velocity dispersions. The spectroscopic redshifts and velocity dispersion measurements along with the combined systematic and statistical errors (Appendix \ref{app:dispersion_test}) are listed in Table \ref{tab:specz_SEDparam} and \ref{tab:galfit}, respectively. In Table \ref{tab:galfit}, we also list the velocity dispersion for UDS19627 derived in \cite{Toft+12}. In Figure \ref{fig:redshift}, the derived spectroscopic redshift are compared with the photometric estimates from \cite{Muzzin_COSMOS_Uvista} and \cite{Laigle+16}. Using the Normalised Median Absolute Deviation ($\sigma_{NMAD}$) from \cite{Brammer+08}, no catastrophic outliers are found except for photometric redshifts being systematically below the spectroscopic redshifts for both catalogs, finding a better agreement for \cite{Muzzin_COSMOS_Uvista}.

\begin{table*}[!t]
\caption{The stellar population model parameters}
\label{tab:specz_SEDparam}
\begin{tabular}{llllllllll}
\hline
Target ID & $z_{\rm{spec}}$  & ${\rm{log}}(M_{\ast}/M_{\odot})$ & $\mathrm{log}(\rm{Age/yr})$ & $\rm{A}({\rm{g}})$ & $\rm{SFR}_{SSP}[M_{\odot}/\rm{yr}]$ &  $\rm{SFR}_{opt}[M_{\odot}/\rm{yr}]$ & $\rm{SFR}_{24}[M_{\odot}/yr]^\mathrm{a}$ \\ \hline
UV-108899               & $2.2312$ & $11.62^{+0.16}_{-0.18}$ & $9.15^{+0.27}_{-0.30}$ & $0.38^{+1.00}_{-0.38}$ & $<13$    &   $6\pm4$ ([OII])        & $<15$                             \\
UV-250513               & $2.0814$ & $11.51^{+0.18}_{-0.19}$ & $9.16^{+0.27}_{-0.31}$ & $0.38^{+1.02}_{-0.38}$ & $<12$    &   $<3$ (H$\alpha$)       & $<13$                             \\
CP-561356               & $2.6963$ & $11.62^{+0.21}_{-0.20}$ & $9.14^{+0.28}_{-0.32}$ & $0.62^{+1.17}_{-0.62}$ & $<86$    &   $<19$ ([OII])          & $<90^{\oplus}$                    \\
UV-105842               & $2.0195$ & $11.68^{+0.16}_{-0.17}$ & $9.19^{+0.26}_{-0.33}$ & $0.81^{+1.04}_{-0.81}$ & $<17$    &   $<2$ (H$\alpha$)       & $19\pm5$                         \\
UV-171687               & $2.1020$ & $11.51^{+0.18}_{-0.19}$ & $9.13^{+0.28}_{-0.32}$ & $0.64^{+1.13}_{-0.64}$ & $<24$    &   $<3$ (H$\alpha$)       & $26\pm6^{\diamond\oplus}$        \\
UV-90676                & $2.4781$ & $11.78^{+0.17}_{-0.18}$ & $9.09^{+0.29}_{-0.29}$ & $0.41^{+0.99}_{-0.41}$ & $<88$    &   $<6$ ([OII])           & $<92^{\diamond\oplus}$            \\
CP-1291751              & $2.0253$ & $11.24^{+0.23}_{-0.22}$ & $9.17^{+0.27}_{-0.33}$ & $0.82^{+1.26}_{-0.82}$ & $<17$    &   $<2$ (H$\alpha$)       & $18\pm4$                         \\
UV-155853               & $1.9816$ & $11.62^{+0.18}_{-0.17}$ & $9.23^{+0.24}_{-0.33}$ & $0.88^{+1.04}_{-0.86}$ & $<14$    &   $<4$ (H$\alpha$)       & $<15$                             \\
UV-171060               & $2.0995$ & $11.48^{+0.16}_{-0.17}$ & $9.16^{+0.27}_{-0.31}$ & $0.41^{+1.03}_{-0.41}$ & $<14$    &   $<2$ (H$\alpha$)       & $<15^{\diamond}$                  \\
UV-230929               & $2.1679$ & $11.48^{+0.16}_{-0.16}$ & $9.10^{+0.28}_{-0.28}$ & $0.22^{+0.89}_{-0.22}$ & $<6$     &   $<4$ ([OII])           & $<7$                              \\
UV-239220               & $2.0057$ & $11.57^{+0.20}_{-0.20}$ & $9.18^{+0.26}_{-0.33}$ & $0.66^{+1.14}_{-0.66}$ & $<19$    &   $35\pm15$ (H$\alpha$)  & $21\pm4^{\diamond\oplus}$        \\
UV-773654               & $2.0328$ & $11.59^{+0.19}_{-0.20}$ & $9.20^{+0.26}_{-0.33}$ & $0.68^{+1.13}_{-0.68}$ & $<12$    &   $<2$ (H$\alpha$)       & $13\pm3^{\diamond\oplus}$        \\
CP-1243752              & $2.0903$ & $11.79^{+0.17}_{-0.17}$ & $9.23^{+0.24}_{-0.32}$ & $0.76^{+1.06}_{-0.76}$ & $<11$    &   $<2$ (H$\alpha$)       & $<12$                             \\
CP-540713               & $2.0409$ & $11.26^{+0.22}_{-0.23}$ & $9.16^{+0.27}_{-0.32}$ & $0.57^{+1.19}_{-0.57}$ & $<10$    &   $<2$ (H$\alpha$)       & $<12$                             \\
UDS-19627$^{\dagger}$   & $2.0389$ & $11.37^{+0.13}_{-0.10}$ & $9.08^{+0.11}_{-0.10}$ & $0.77^{+0.36}_{-0.32}$ & ...     &   $<6$ (H$\alpha$)       & $<40^\mathrm{b}$                  \\ \hline
\end{tabular}
\tablecomments{Column 1: Target ID, Column 2: Spectroscopic redshift, Column 3: Stellar mass, Column 4: mass-weighted stellar age, Column 5: Extinction in \textit{g}-band, Column 6: $3\sigma$ upper limit percentiles (representing $99.73\%$ Gaussian confidence intervals) of the stellar population modeled SFR/100 Myr distribution, Column 7: $3\sigma$ SFR upper limits based on H$\alpha$ or $[\rm{OII}]\lambda3727$ (Section \ref{sec:SFR_uplim}), Column 8: $24\ \mu{m}$ estimated SFR (see Section \ref{sec:mips24_sfr}).}
\tablenotetext{\dagger}{The values listed for UDS19627 is from \cite{Toft+12}. From this study the $A(v)$ extinction instead of the listed $A(g)$ is qouted.}
\tablenotetext{a}{Galaxies with detections in $1.4$ GHz ($^{\diamond}$) and $3$ GHz ($^{\oplus}$) are indicated with matching symbols.}
\tablenotetext{b}{$2\sigma$ $24\ \mu{m}$ SFR upper limit using method from \cite{Franx+08}} 
\end{table*}

\subsection{Emission lines} \label{sec:emission}
No on-source nebular line emission is detected at $3\sigma$ for any objects in the sample. For UV-108899 and UV-239220 we find indications of emission ($\sim2\sigma$) from [OII]$3726.2,3728.9$ and H$\alpha6563$, respectively. In Appendix \ref{app:on_source_emission}, we discuss the specifics of the fitting method and list, in Table \ref{tab:specz_SEDparam}, the SFR and uncertainties from the [OII] and H$\alpha$ \citep{Kennicutt1998}. Furthermore, spatially offset line emission is observed in four (UV-155853, UV-171687, UV-171060, UV-105842) 2-D spectra coinciding with close proximity sources. In 3/4 cases, this emission arises from foreground or background sources (Appendix \ref{sec:foreground_background}). The latter source north-east of UV-105842 shows significant [OII]$3726.2,3728.9$ \AA{}, [OIII]$4959,5007$\AA{}, and H$\alpha$ emission with a matching redshift of $z=2.0124$. This corresponds to a velocity offset of $2130\pm120$ km/s from UV-105842. If purely due to galaxy motion, such an offset suggests that the two sources are not gravitationally bound at the time of observation. Another explanation of the asymmetric morphology might be a high redshift analog of the locally observed offset AGN \citep{Comerford&Greene+14}, likely caused by recent merger event.

\subsection{Stellar population modeling of continuum emission} \label{sec:stelpop}
In order to put constraints on the physical parameters of the stellar populations, the VIS+NIR X-shooter spectra and the broadband photometry are fit with the Bayesian approach, from \cite{Gallazzi+05} (recently revised in \cite{Zibetti+17}), using the derived spectroscopic redshift. Spectral regions of poor atmospheric transitions are not included in the calculation. Before fitting, the models are convolved by the initial velocity dispersion estimated in Section \ref{sec:veldisp}.

Models are obtained by convolving the latest revision of BC03 Simple Stellar Population (SSP) models using the MILES stellar libraries \citep{Sanchez-Blazquez+06,Falcon-Barroso+11} with a large Monte Carlo library of star formation histories, metal enrichment histories and dust attenuations. The prior distribution of models is the one described in \cite{Zibetti+17}, but here limited to 50,000 models with formation ages younger than 5 Gyr to be consistent with the high redshift of our galaxies. A full description of the model library is given in \cite{Zibetti+17}, however the most relevant information are summarized here.

SFHs are modeled with a continuous component parametrized \`{a} la \cite{Sandage+86}\footnote{$SFR(t) = t/\tau \times exp(-t^2/(2\tau^2))$, see e.g. Section 3.1 in \cite{Zibetti+19}}, thus allowing for both an increasing and a decreasing SFH phase, on top of which random bursts of star formation are added. Stellar metallicity evolves according to the SFH (see \cite{Zibetti+19}), with initial and final values randomly generated in the range $1/50-2.5 Z_\odot$. Finally, for 75\% of the models, the effect of dust attenuation is included following the model of \cite{Charlot_Fall+00} that separates the contribution of the birth clouds affecting stars younger than $10^7$ yr and the contribution of the ISM affecting stars of all ages.

The Bayesian modeling approach assumes the likelihood of each model to be $\propto exp(-\chi^2/2)$. The probability distribution function (PDF) of each physical parameter of interest are computed by weighing the prior distribution of the models in a given parameter by their likelihood, marginalizing over all the other parameters. We additionally used the information from the mid-IR flux limit to restrict the sample of acceptable models to those that have a SFR consistent with the $24\ \mu{m}$-based upper limits and detections (see Section \ref{sec:mips24_sfr}). The median and the $16^{th}$ and $84^{th}$ percentiles of the PDFs are adopted as the fiducial estimates and their uncertainties for each parameter. Note that this approach allows the derivation of realistic uncertainties on the key physical parameters, accounting for both the observational errors and the intrinsic degeneracies among different parameters.\\

The stellar mass, mass-weighted mean stellar age, effective dust attenuation ($A(g)$) and SFR, averaged over the last $100$ Myr for our sample, are reported in Table \ref{tab:specz_SEDparam}. In this table, the SFR limits from nebular line and $24\ \mu{m}$ emission (see Sections \ref{sec:SFR_uplim} and \ref{sec:mips24_sfr}) are also listed. Stellar masses are within the range of $\mathrm{log}_{10}(M_{\ast}/M_{\odot}) = 11.23-11.79$, with a median of $11.57$. Compared to \cite{Belli+17}, this sample is on average more massive, which is reflected by the brighter K-band magnitudes (see Figure \ref{fig:selection}). Such massive quiescent galaxies have also been found over a larger area in \cite{Arcila-Osejo+19}. The SFR limits and dust-corrected stellar masses, together with the mean stellar mass weighted ages of $\sim1.4$ Gyr, confirm the expectations from the selection that this is, in fact, a sample of massive recently quenched galaxies. Three of the galaxies are double sources and the stellar masses are corrected in Section \ref{sec:major_merger_spectra}.

\subsection{Star formation and quiescence} \label{sec:SFR_quiescence}

\subsubsection{Rest-frame optical emission lines} \label{sec:SFR_uplim}

In order to confirm the quiescence nature of our galaxies upper limits on $[\rm{OII}]\lambda3727$ and H$\alpha$ emission are measured. These are converted into upper limits of the unobscured SFRs following Equation (2) and (3) in \cite{Kennicutt1998}, under the assumptions of solar abundance ratio and that all massive star formation is traced by ionized gas. A $3\sigma$ flux upper limit is determined by summing up the flux error density squared over a region of $\Delta{\lambda}=1000$ km/s (similar to $300-500$ km/s line dispersions):
\begin{equation} F_{3\sigma\ limit} = 3\sqrt{\sum{\sigma_{flux}^2\delta{\lambda}^2}}. \end{equation}
Here $\sigma_{flux}$ and $\delta{\lambda}$ are the flux uncertainty and bin size, respectively. Note that we do not introduce any dust extinction in this conversion, as this is largely unconstrained (see Section \ref{sec:SFR_tracers} for an estimated upper limit on the dust extinction). We find unobscured SFR upper limits that are consistent with the expectation that these galaxies are quiescent ($-10<{\rm{log}}_{10}(\rm{sSFR/yr})<-11.5$). The difference between the $[\rm{OII}]\lambda3727$ and H$\alpha$ SFR limits are $<11\ M_{\ast}/\rm{yr}$, and in Table \ref{tab:specz_SEDparam} the lowest SFR upper limits are listed.

\begin{figure}
\centering
  \includegraphics[width=8.5cm]{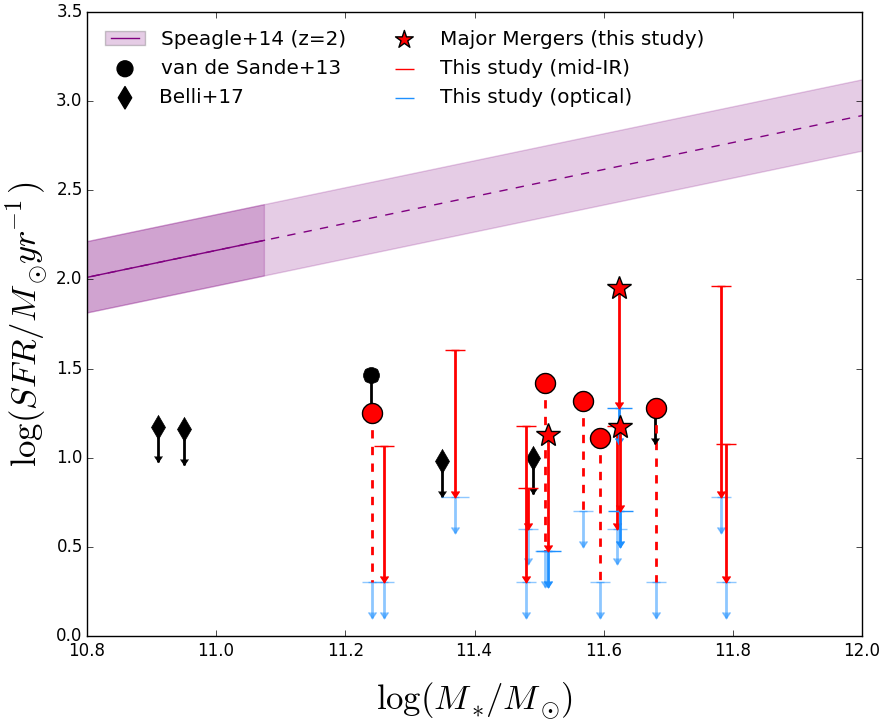}
  \caption{SFR -- $M_{\ast}$ plane for massive quiescent galaxies at $z>2$ with $24\ \mu{m}$ coverage. The SFR main-sequence at $z=2$ from \cite{Speagle+14} is shown in dark purple, with its 0.2 dex ($1\sigma$) scatter. The light purple region extending beyond $\mathrm{log}(M_{\ast}/M_{\odot})>11.1$ is an extrapolation of the best-fit relation. The $24\ \mu{m}$ MIPS SFR detections (red circles)/upper limits (red arrows) are shown, with the major mergers (composite measurement of the SFR) in red stars. We show our rest-frame optical $\rm{SFR}_{3\sigma}$ (based on [OII] and H$\alpha$) in blue upper limits.  We show $24\ \mu{m}$ SFR upper limits for the $2$ objects from \cite{vandeSande2013} (circles), together with $4$ dust-corrected H$\alpha$ upper limits from \cite{Belli+18} (diamonds) in black upper limits. Our sample of galaxies have suppressed SFR compared to the main-sequence at $z=2$ and can be considered truly quiescent galaxies.} 
  \label{fig:SFR_Mstar}
\end{figure}

\subsubsection{Mid-infrared emission} \label{sec:mips24_sfr}
The SFR, derived from rest-frame optical emission lines, represents a lower limit to the total star formation in the presence of strong dust attenuation. Therefore, the SFR from the \textit{Spitzer}/MIPS $24\,\mu{m}$ emission \citep{Wu+05,Zhu+08,Rieke+09,Kennicutt+09} are estimated under the assumption of zero or subdominant AGN emission. Here, the $24\,\mu{m}$ flux densities (or 3$\sigma$ upper limits for sources undetected at 24$\,\mu{m}$), from the most recent  ``super deblended''  FIR COSMOS catalog \citep{Jin+18}, are adopted.
To derive SFR estimates, the $z=2$ main-sequence SED template of \cite{Magdis+12} is rescaled to the measured $24\,\mu{m}$ flux densities (or the  3$\sigma$ upper limits) of our targets. The emerging total infrared luminosity ($L_{\rm IR}$) of the templates are converted to SFR through the $L_{\rm IR}$-SFR relation of \cite{Kennicutt1998}, tuned to the adopted Chabrier IMF of this study.
Detections corresponding to a median ${\rm{SFR}}\sim20\ M_{\odot}\rm{yr}^{-1}$ are found for $5$ of the galaxies that are undetected in the $24\ \mu{m}$ catalog \citep{Le_Floch+09}.
The remaining galaxies are not individually detected and we thus fix them to their $3\sigma$ upper limit. UV-90676 and CP-561356 that have upper limits of $\lesssim{}90\ M_{\odot}\rm{yr}^{-1}$. Both galaxies show strong merger signatures (see Section \ref{sec:major_merger_spectra}). The derived $24\ \mu$m SFR are listed in Table \ref{tab:specz_SEDparam}.

\subsubsection{Comparison of different star formation tracers} \label{sec:SFR_tracers}
Figure \ref{fig:SFR_Mstar} shows the position of the sample of MQGs in the ${\rm{log}}(\rm{SFR})-{\rm{log}}(M_{\ast})$ main-sequence at $z=2$. For reference, the SFR main-sequence at matching redshift from \cite{Speagle+14} is shown, extrapolated to the stellar mass range ${\rm{log}}_{10}(M_{\ast}/M_{\odot})>11.1$ covered by our galaxies. 

The rest-frame optical SFR limits are systematically lower than the mid-IR estimates (both probing $10-100$ Myrs timescales). This suggests either that the star-forming regions are strongly obscured and/or AGN dust heating \citep{Fumagalli+14}. Under the assumption of no AGN contribution to the heating that produces the mid-IR emission (see also Section \ref{sec:disc_agn}), the dust extinction is estimated by comparing the obscured and un-obscured SFR estimates, resulting in a mean extinction of $A(v)<1-2$ consistent with our SED fit derived $A(g)$ (g-band) extinction. In order to judge if a significant contribution to the mid-IR heating arises from AGN, we check if there are any radio counterparts detected in \cite{Jin+18}. Radio emission is detected in 5 sources at $1.4$ GHz and in 5 sources at $3$ GHz (indicated with symbols in Table \ref{tab:specz_SEDparam}), showing that AGN heating could be responsible for the elevated mid-IR SFR estimates. Further treatment of the radio detections will be part of a future paper (Cortzen at al. in prep).

The SFRs derived from our stellar population analysis (Section \ref{sec:stelpop}) are consistent with $\rm{SFR}\sim{}0\ M_{\ast}\rm{yr}^{-1}$ for all galaxies in our sample. In Table \ref{tab:spec_info}, we list the $3\sigma$ upper limits on these SFR limits. However, even considering the most conservative upper limits on the SFR from the 24 $\mu{m}$ emission, our sample of MQGs lies $\sim2$ dex below the SFR main-sequence at their redshifts, confirming their quiescent nature.\\
\mbox{}\\

\begin{figure*}[!htbp]
\centering
      \includegraphics[width=14.5cm]{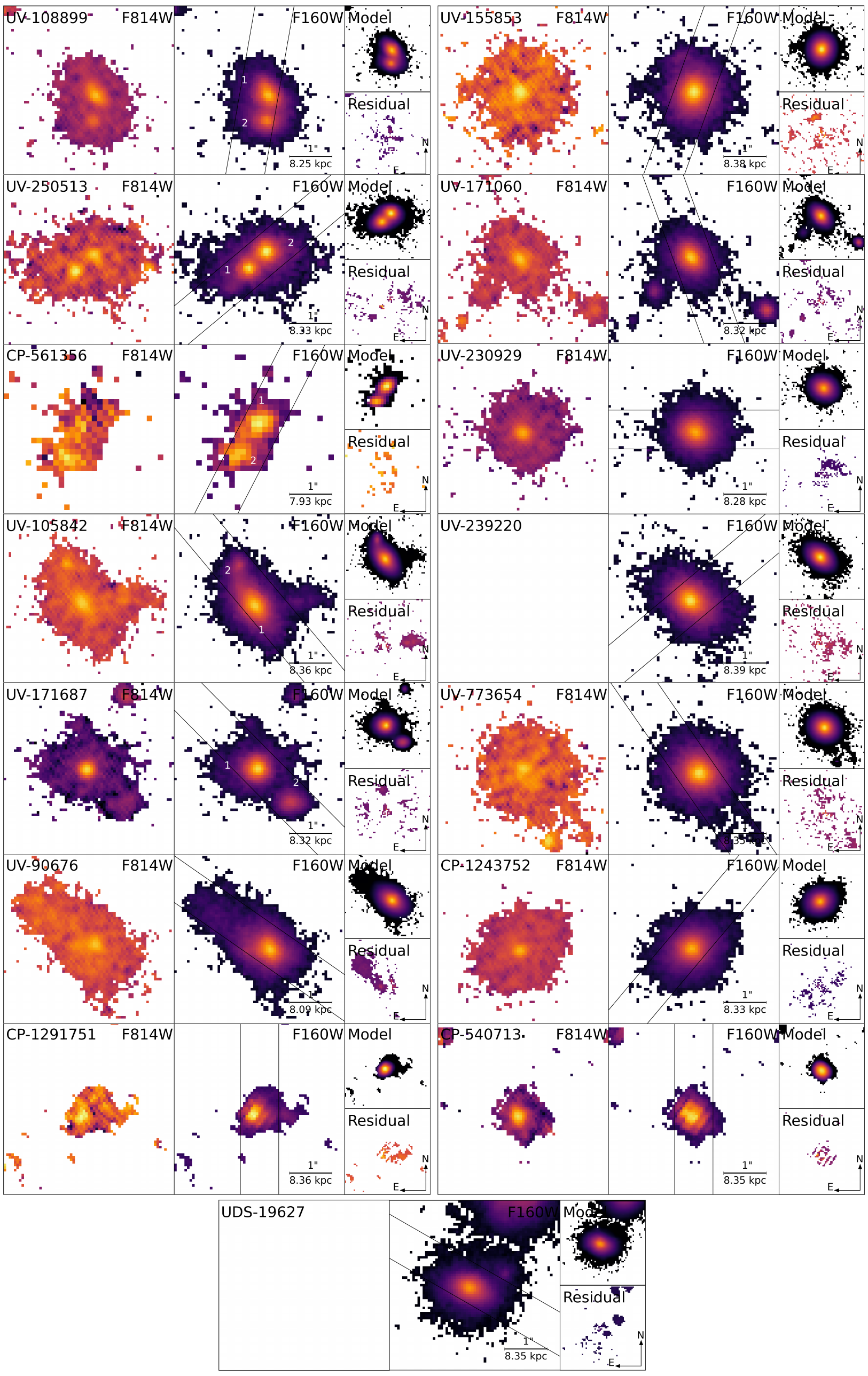}
      \caption{$I_{F814W}$, $H_{F160W}$, \textit{GALFIT} model, and \textit{GALFIT} residual for our sample of massive quiescent galaxies at $z>2$ in $4x4"$ cutouts. Pixels, with a $3\sigma$ confidence (w.r.t. background), are indicated with a logarithmic color scale to showcase the structure and morphology of the sample. $H_{F160W}$ significant pixels are used as a mask for all the images. In the residual image, the pixels, one standard deviation above the background, are shown within this mask. The X-shooter slit is overlaid at the orientation of the spectroscopic observations. A scale of $1\farcs0$ is shown in kpc for size reference.}
  \label{fig:HST_presentation}
\end{figure*}

\begin{table*}[!t]
\caption{Summary of structural properties}
\label{tab:galfit}
\begin{tabular}{llllllllccl}
\hline
Target ID   & $z_{\rm{phot}}$ & $z_{\rm{spec}}$        & $\sigma$ [km/s]    & $R_{maj}$ [kpc]   & $n$      & $q$     & $R_{\rm{Flux}}^\mathrm{b}$ & ${\rm{log}}_{10}(M_{\ast,c}/M_{\odot})$ & ${\rm{log}}_{10}(M_{\rm{dyn}}/M_{\odot})$ & Class$^\mathrm{c}$ \\ \hline
UV-108899-1$^\mathrm{a}$ & $2.19$  & $2.2312$            & $470\pm82$  & $1.36\pm0.14$   & $2.51$ & $0.44$ & $ 0.56$    & $11.38^{+0.16}_{-0.18}$   & ...              & P$^{\dagger}$ \\
UV-108899-2$^\mathrm{a}$ & ...     & ...                 & ...         & $3.38\pm0.34$   & $7.15$ & $0.56$ & $0.44$     & $11.26^{+0.17}_{-0.17}$   & ...              & P$^{\dagger}$ \\
UV-250513-1$^\mathrm{a}$ & $2.03$  & $2.0814$            & $174\pm44$  & $3.84\pm0.38$   & $4.00$ & $0.59$ & $0.55$     & $11.26^{+0.17}_{-0.19}$   & ...              & P$^{\dagger}$ \\
UV-250513-2$^\mathrm{a}$ & ...     & ...                 & ...         & $1.60\pm0.16$   & $4.00$ & $0.63$ & $0.45$     & $11.16^{+0.17}_{-0.18}$   & ...              & P$^{\dagger}$ \\
CP-561356-1$^\mathrm{a}$ & $2.43$  & $2.6963$            & $280\pm128$ & $4.14\pm0.41$   & $1.45$ & $0.62$ & $0.71$     & $11.47^{+0.21}_{-0.20}$   & ...              & P$^{\dagger}$ \\
CP-561356-2$^\mathrm{a}$ & ...     & ...                 & ...         & $2.78\pm0.28$   & $0.90$ & $0.39$ & $0.29$     & $11.09^{+0.21}_{-0.21}$   & ...              & P$^{\dagger}$ \\
UV-105842-1              & $1.93$  & $2.0195$            & $263\pm57$             & $4.07\pm0.41$   & $3.51$ & $0.51$ & $1.00$     & $11.68^{+0.16}_{-0.17}$   & $11.61\pm0.19$   & P             \\
UV-171687-1              & $2.04$  & $2.1020$            & $182\pm50$             & $5.12\pm0.51$   & $4.00$ & $0.77$ & $1.00$     & $11.51^{+0.18}_{-0.19}$   & $11.37\pm0.24$   & P             \\
UV-90676                 & $2.57$  & $2.4781$            & $347\pm82$             & $5.22\pm0.51$   & $4.98$ & $0.61$ & $1.00$     & $11.78^{+0.17}_{-0.18}$   & $11.89\pm0.21$   & P             \\
CP-1291751               & $2.06$  & $2.0253$            & ...                    & $3.47\pm0.35$   & $3.59$ & $0.67$ & $1.00$     & $11.24^{+0.23}_{-0.22}$   & ...              & P             \\
UV-155853                & $1.96$  & $1.9816$            & $247\pm30$             & $4.55\pm0.46$   & $3.62$ & $0.85$ & $1.00$     & $11.62^{+0.18}_{-0.17}$   & $11.60\pm0.11$   & E             \\
UV-171060                & $2.02$  & $2.0995$            & ...                    & $1.73\pm0.17$   & $4.00$ & $0.54$ & $1.00$     & $11.48^{+0.16}_{-0.17}$   & ...              & E             \\
UV-230929                & $2.09$  & $2.1679$            & $252\pm21$             & $1.74\pm0.17$   & $3.01$ & $0.73$ & $1.00$     & $11.48^{+0.16}_{-0.15}$   & $11.23\pm0.08$   & E             \\
UV-239220                & $2.00$  & $2.0057$            & ...                    & $5.35\pm0.54$   & $4.21$ & $0.62$ & $1.00$     & $11.57^{+0.20}_{-0.20}$   & ...              & E             \\
UV-773654                & $1.96$  & $2.0328$            & ...                    & $3.77\pm0.38$   & $3.34$ & $0.84$ & $1.00$     & $11.59^{+0.19}_{-0.19}$   & ...              & E             \\
CP-1243752               & $2.01$  & $2.0903$            & $350\pm53$             & $2.85\pm0.29$   & $4.50$ & $0.79$ & $1.00$     & $11.79^{+0.17}_{-0.17}$   & $11.66\pm0.14$   & E             \\
CP-540713                & $1.98$  & $2.0409$            & $353\pm97$             & $1.65\pm0.17$   & $0.96$ & $0.79$ & $1.00$     & $11.26^{+0.22}_{-0.23}$   & $11.59\pm0.24$   & E             \\
UDS-19627                & $2.02$  & $2.0389$            & $318\pm53$             & $2.00\pm0.20$   & $3.32$ & $0.51$ & $1.00$     & $11.37^{+0.13}_{-0.10}$   & $11.48\pm0.15$   & E             \\ \hline
\end{tabular}

\tablecomments{Column 1: Target ID, Column 2: Photometric redshift from \cite{Muzzin_COSMOS_Uvista}, Column 3: Spectroscopic redshift (Section \ref{sec:veldisp}), Column 4: Stellar velocity dispersion measurement (Section \ref{sec:veldisp}), Column 5: Effective semi-major axis (Section \ref{sec:galfit_modeling}), Column 6: S\'ersic index \citep{Sersic}, Column 7: Axis ratio $q=b/a$ (defined by the ratio between the semi minor and major axis), Column 8: Flux scaling used to estimate the corrected stellar mass, ${\rm{log}}_{10}M_{\ast,c}$ (Section \ref{sec:major_merger_spectra}), Column 9: Corrected stellar mass (Section \ref{sec:major_merger_spectra}), Column 10: Dynamical mass (Section \ref{sec:dynamicalmass}), Column 11: Morphological classification (P:Peculiar, E:Elliptical) from \cite{Conselice+05}.}

\tablenotetext{\dagger}{Galaxies classified as Major Mergers in Section \ref{sec:major_merger_spectra}} 
\tablenotetext{a}{Double sources have similar photometric and spectroscopic redshift as well as the stellar velocity dispersion estimated from their composite spectrum}
\tablenotetext{b}{Relative Flux Ratio = $F_{i}/(F_{i}+F_{j})$} 
\tablenotetext{c}{Galaxies marked with ($\dagger{}$) are classified as major mergers in Section \ref{sec:major_merger_spectra}}

\end{table*}

\subsection{Galaxy structure and sizes} \label{sec:galfit_modeling}
The 2-D stellar light distribution traced by \textit{HST}/WFC3 $H_{F160W}$ imaging are modeled with the $\chi^2$-minimization fitting code \textit{GALFIT} \citep{Peng_Galfit} in order to retrieve the structural parameters of our sample of MQGs. A first run of \textit{SExtractor} \citep{SExtractor} allows us to detect the objects in each field and to obtain an initial guess for the structural parameters. Postage stamp for each target is constructed such that it encloses an ellipse with a major axis 2.5 times the Kron radius obtained by \textit{SExtractor}. The local sky level in each stamp is calculated using Galapagos \citep{Galapagos}. This sky level is passed to \textit{GALFIT} and kept fixed during the fitting. For the WFC3 data, a combination of the TINYTIM\footnote{\url{http://www.stsci.edu/hst/observatory/focus/TinyTim}}-simulated point spread function (PSF) and an empirical stacked star PSF are used. For the NICMOS data, an empirical stacked PSF are used.\\

\begin{figure}[!t]
\centering
      \includegraphics[width=8.5cm]{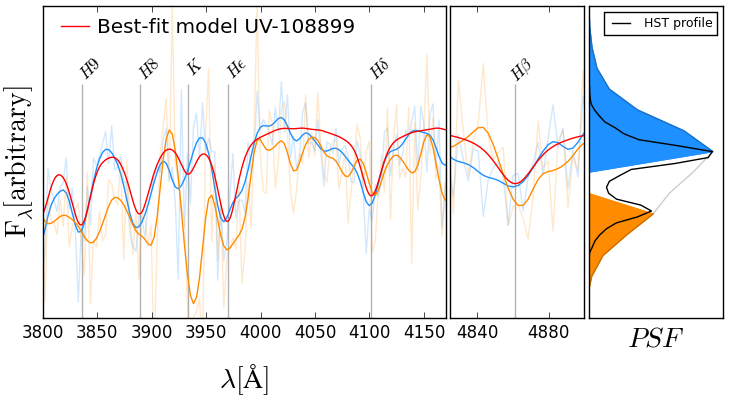}
      \includegraphics[width=8.5cm]{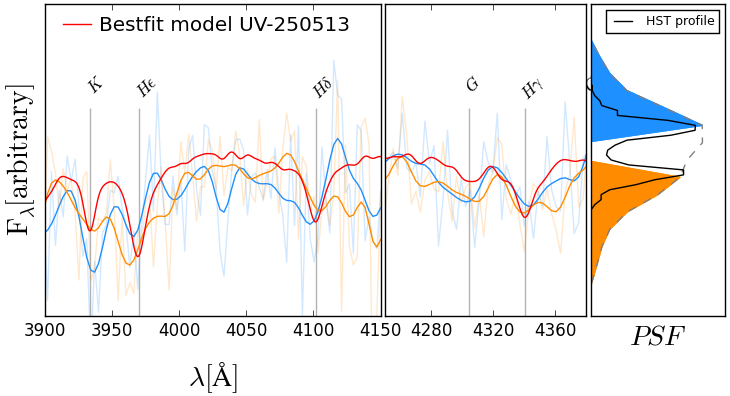}
      \includegraphics[width=8.5cm]{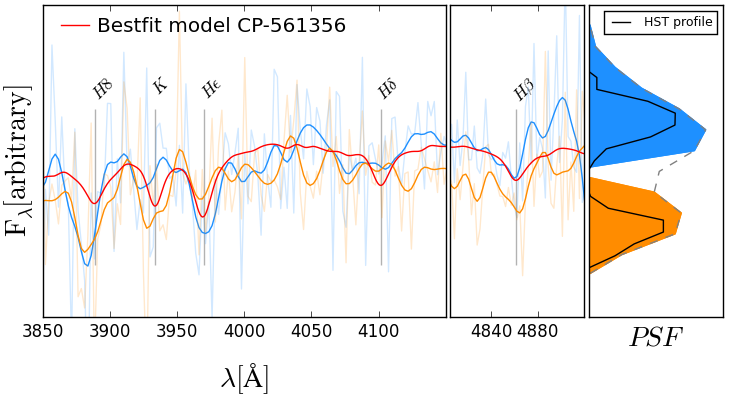}
      \caption{Individual flux extractions (blue, orange) from spatially divided 2-D spectra of the major merger candidate sources UV-108899, UV-250513, CP-561356 (top to bottom). The thick line is a smoothed version of the original spectra shown by the thin line. The right panel shows the wavelength collapsed 2-D spectrum (grey line) color coded to match the individual extracted 1-D spectra (left, center panel). For reference the 1-D resolved \textit{HST} $H_{F160W}$ profile is shown in thin black line. The best fit model of the composite spectrum is shown in red and the visible Balmer absorption lines are indicated. For each galaxy we confirm the spectroscopic redshift proximity by the matching of absorption lines and conclude that these sources are ongoing major mergers.}
  \label{fig:mm_sep}
\end{figure}

Finally, \textit{GALFIT} is run on each postage stamp, adopting a flexible S\'ersic profile for every source \citep{Sersic},
\begin{equation}\label{eq_sersic}
 \Sigma(R)=\Sigma_{e}\exp\left\{-\kappa_n\left[\left(\frac{R}{R_e}\right)^{1/n}-1\right]\right\}.
\end{equation}
The parameter $R_e$ is the effective radius enclosing half of the flux from the model light profile, $\Sigma(R_{e})$ is the surface brightness at the effective radius and $n$ is the S\'ersic index. The quantity $\kappa_n$ is a function of the S\'ersic index, which defines the global curvature of the light profile, and is obtained by solving the equation $\Gamma(2n)=2\gamma(2n,\kappa_n)$, where $\Gamma$ and $\gamma$ are, respectively, the gamma function and the incomplete gamma function.

\textit{GALFIT} is run several times to ensure that the solutions correspond to a global minimum in the minimization algorithm for each image, by varying the initial guesses of the total magnitude, effective radius and S\'ersic index. The parameters are constrained so to avoid any unphysical solutions (effective radius $>0.2$ pixels, $q > 0.1$, $0.5<n < 8$). Initially, all targets are fit with $n$ as a free parameter. In unstable cases where the maximum or minimum $n$ are reached, the images fixing the S\'ersic index at either $n=1$ or $n=4$ are re-fit, choosing the model providing the smallest $\chi^2$ as the best-fit solution. These two choices represent realistic descriptions of an early-type galaxy dominated by either a disk or a bulge. Throughout the whole fitting procedure, neighboring objects are either modeled or masked, depending on their proximity to the main target. A $10\%$ measurement uncertainty on the size is \citep{vanderWel+08,Newman+12} shown to be a fair representation. This conservative error estimate is thus adopted. The semi-major axis, $R_{\rm{e,maj}}$, is adopted as the effective radius in the following sections. The best-fit parameters and their uncertainties are reported in Table \ref{tab:galfit}.\\

In Figure \ref{fig:HST_presentation}, we present the rest-frame UV ($I_{F814W}$) and optical ($H_{F160W}$) images along with the \textit{GALFIT} model and residual. The morphologies of these galaxies are classified in the $H_{F160W}$ image according to \cite{Conselice+05} and they fall into the two categories for quiescent systems: Ellipticals (E) and Peculiars (P). When available, the spectroscopic observations are used to determine the distance in redshift space to objects that fall in the X-shooter slit (see Section \ref{sec:emission}).  The majority of sources turn out not to be associated with the central galaxy. $9/15$ galaxies are categorized as Elliptical galaxies while the remaining are categorized as Peculiar galaxies with major mergers (UV-108899, UV-250513, CP-561356), minor mergers (UV-105842, CP-1291751) and/or strong tidal/post-merger features (UV-105842, UV-90676). The galaxies UV-108899, UV-250513 and CP-561356 are confirmed as ongoing major mergers in the following section. The classifications and the morphological parameters are listed in Table \ref{tab:galfit}.

\subsection{Spectroscopic confirmation and stellar mass correction of ongoing major mergers} \label{sec:major_merger_spectra}

The RB color images, in Figure \ref{fig:spec_presentation}, reveal that three galaxies (UV-108899, UV-250513, CP-561356) appear to be double systems. The spectra, shown in the same figure, are the total extraction of the combined light from the two galaxies. These objects are within close proximity and the light in the reduced 2-D frames are blended to an unknown extent (due to limited seeing). At the expense of drastically decreasing the S/N, an attempt to separate the sources and determine if their individual redshift measurements can confirm their proximity are made.

For each system, the resolved 1-D \textit{HST} $H_{F160W}$ light profile (extracted parallel to the X-shooter slit) is overlaid on top of the wavelength collapsed 2-D spectrum trace. A double Gaussian profile fit allowed us to gauge the amount of blending and to make a conservative extraction of each individual galaxy, minimizing cross-source contamination. In Figure \ref{fig:mm_sep}, the individual extractions and the best-fit to the composite spectrum from Section \ref{sec:stelpop} are shown.

Because of the low S/N of the individual conservative flux extractions, the estimation of the velocity offset are refrained, since it would be dominated by large uncertainties. However, the galaxies are within close physical proximity due to the matching absorption lines shown in the figure and can be considered ongoing quiescent (dry) major-mergers. This confirmation is important as, in the following section, it can be used to correct their stellar masses, prior to presenting them in the mass-size plane (see Section \ref{sec:mass_size}).\\

Spectroscopic confirmation allows us to deblend the composite stellar mass of each system using the $H_{F160W}$ magnitude as a proxy for tracing the bulk of the stars in the galaxies. The \textit{GALFIT} modeled $H_{F160W}$ flux ratio supports the fact that these galaxies are major mergers with mass ratios of $1:1-3$. We used the flux ratio to correct the stellar masses as:
\begin{equation} \label{eq:stellar_mass_corr}
M_{\ast,i} = M_{\ast,tot}\frac{F_i}{F_i+F_j} =  M_{\ast,tot} R_{\rm{Flux}},
\end{equation}
where $i$ and $j$ refer to the two merging galaxies and $F$ is the total flux from \textit{GALFIT}. The corrected stellar masses ($M_{\ast,c}$) and the relative flux ratio scaling, $R_{\rm{Flux}}$, are listed in Table \ref{tab:galfit}, with sources names matching the numbering in Figure \ref{fig:HST_presentation}. Following this correction, the galaxies still classify as MQGs with stellar masses, ${\rm{log}}_{10}(M_{\ast}/M_{\odot})>11$.


\section{Results} \label{sec:results}

\subsection{Minimal progenitor bias} \label{sec:minimal_progenitor_bias}

A major issue preventing us from deriving a consistent evolutionary picture connecting galaxy populations across time is the ``progenitor bias'' problem \citep[e.g.][]{vanDokkum_Franx+96,Carollo+13}. When comparing galaxies across time, the implicit assumption is that the high redshift sample contains all progenitors of the low redshift reference sample. However, the fraction of quenched galaxies has been found to grow over time \citep{Buitrago+13} introducing an unknown bias when comparing samples of galaxies across different epochs. 

One approach, that has been suggested to minimize the progenitor bias, is comparing the evolution of galaxies at fixed velocity dispersion \citep[see e.g.][]{Belli+14a}. Archaeological studies \citep{vanderWel+09a,Graves+09,Bezanson+12} find evidence suggesting that the velocity dispersion in quiescent galaxies remains approximately unchanged across cosmic time ($z<1.5$). In such a scenario the velocity dispersion must be weakly affected by the average merger history, which according to the numerical study by \cite{Hilz+12} occurs for minor merger-driven evolution. A detailed discussion on fixed velocity dispersion evolution is given in \cite{Belli+14a,Belli+17}. Another way to minimize the progenitor bias has been to study galaxy populations at constant cumulative number density (CND) instead of fixed velocity dispersion or stellar mass \citep[see e.g.][]{Mundy+15}. This approach are introduced in \cite{vanDokkum+10} and refined further in \cite{Behroozi+13} and \cite{Leja+13}. 
In Section \ref{sec:massive_n}, a sample of massive galaxies with central stellar population ages suggesting formation at $z>2$ are introduced. This sample is volume limited and represents the most massive early-type systems observed in the local Universe. In order to draw a meaningful comparison, a subgroup of the most massive galaxies at $z=0$ are selected and matched with the CND at $z=2$. This will now be referred to as the ``fixed'' CND. This approach is based on the assumption that the rank of galaxies, within the stellar mass function, is not strongly affected across cosmic time. This occurs if the stellar mass continuously grows from $z=2-0$, implying the availability of surrounding material to accrete (or events that trigger secondary SF, although this is not expected for the massive quiescent galaxies at $z>2$)
\citep{Brammer+11,Behroozi+13,Muzzin13b,Marchesini+14}. 

First, the CND of massive (${\rm{log}}(M_{\ast}/M_{\odot})>11.2$) \textit{UVJ} quiescent galaxies in the redshift range $1.9<z<2.5$ is estimated using the \cite{Muzzin_COSMOS_Uvista} catalog. The stellar-mass limit represents the lower limit on the standard deviation of the mean stellar mass from the sample of galaxies studied in this paper. Our sample is $22\ \%$ stellar mass complete using these selection criteria. We count $58$ galaxies inside a comoving volume spanned by this redshift range giving a $n ({\rm{log}}(M_{\ast}/M_{\odot})>11.2) = 9.7\times10^{-6}\ Mpc^{-3}$.

The MASSIVE galaxy sample is trimmed starting from the most massive object of the survey and including progressively less massive systems until we reach the fixed CND of the massive \textit{UVJ} quiescent galaxies at $z\sim2$. The final fixed CND-matched MASSIVE sample consists of the $25$ most massive local elliptical galaxies with stellar masses of ${\rm{log}}(M_{\ast}/M_{\odot})>11.70$. The fixed CND-matched MASSIVE sample is referred to as ``MASSIVE(n)'' hereafter. The MASSIVE(n) sample is considered a minimal progenitor biased sample and used as our local reference sample in Section \ref{sec:dispersion_size}, \ref{sec:mass_size}, and \ref{sec:dynamicalmass}.\\

The CND evolution suffers from large uncertainties from individual merger histories causing scatter in the mass rank which is the main uncertainty for the highest stellar masses \citep{Behroozi+13,Torrey+17}. In \cite{Torrey+17}, they estimate the mass rank scatter for ${\rm{log}}_{10}(M_{\ast}/M_{\odot})>11$ galaxies in Illustris \citep{Genel+14_Illustris,Nelson+15} by forward modeling of the cumulative number density. Their forward modeling, referred to as Density Distribution Functions, is well described by a lognormal distribution and the uncertainties can thus be treated as confidence intervals. For massive galaxies the dominating uncertainty, the mass rank scatter, introduces a uncertainty of factor of $\sim2$ (within $80\ \%$ confidence intervals) on the CND following the evolution from $z=2$ to $0$. In \cite{Behroozi+13}, they find a similar uncertainty for the fixed CND evolution. This uncertainty on the CND evolution from the mass rank scatter is adopted and used to repeat the selection of the local reference sample resulting in a corresponding uncertainty on the limit of the stellar mass cut ${\rm{log}}(M_{\ast}/M_{\odot})>11.70^{+0.07}_{-0.10}$ and thus the number of galaxies in the local reference sample.

As an alternative approach to the fixed CND matching, the probabilistic approach from \cite{Wellons&Torrey+17} is used to estimate the CND at $z=0$. In Appendix \ref{app:CND}, the results (from Figure \ref{fig:dispersion_radii}, \ref{fig:masssize} and \ref{fig:Mstar_Mdyn}) for both a fixed and probabilistic CND matching approach is presented. The choice of CND-matching method does not affect the qualitative results of this paper.

\subsection{Kinematic evolution of massive quiescent galaxies from $z=2$ to $0$} \label{sec:dispersion_size}
\begin{figure}
  \centering
  \includegraphics[width=8.5cm]{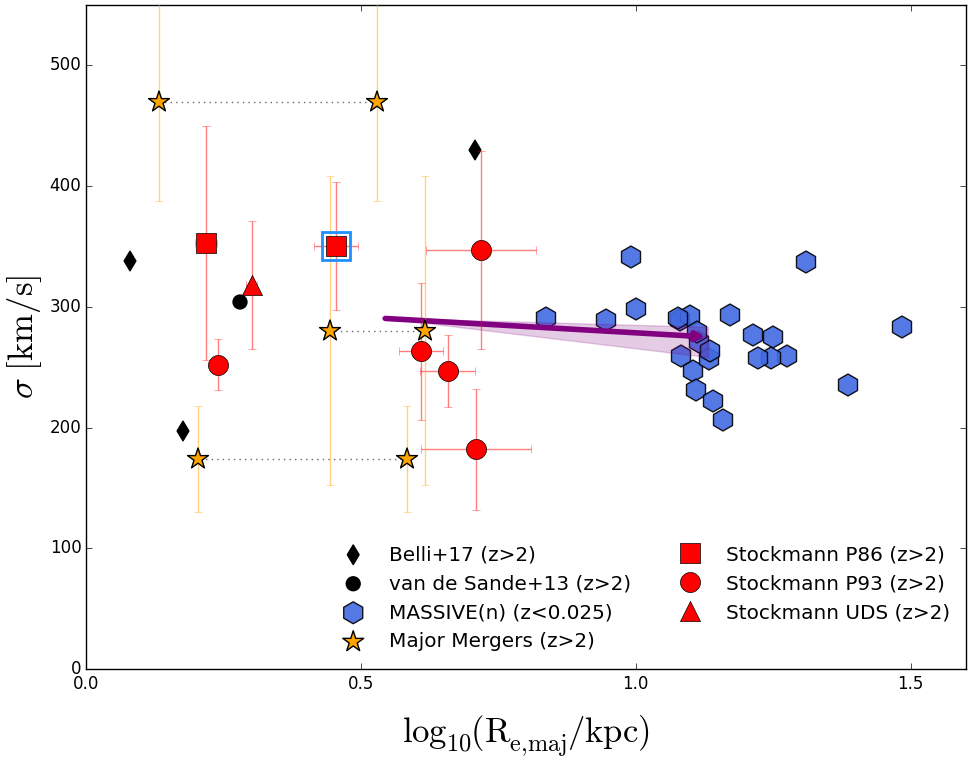}
    \caption{The velocity dispersions are plotted with effective radii for three samples; 1) Our sample (red symbols like Figure \ref{fig:selection}), 2) massive, ${\rm{log}}_{10}(M_{\ast}/M_{\odot})>11$, other quiescent galaxies at $z_{\rm{spec}}>2$ \citep{vandeSande2013,Belli+17}, 3) the MASSIVE(n) sample in blue hexagons. The composite dispersion measurements of the major-merger galaxies are shown in orange stars connecting their individual size measurements with a horizontal dotted line. The blue square indicates our source CP-1243752 \citep[recently published in][]{vandeSande2013,Kriek+16,Belli+17}. The purple arrow shows the median evolution between our study and the MASSIVE(n) sample. The uncertainty from the mass rank scatter on the fixed CND is shown in purple shading. The median evolution between the our study and the MASSIVE(n) sample show evidence for shallow or no kinematic evolution from $z=2$ to $0$.}
  \label{fig:dispersion_radii}
\end{figure}

In Figure \ref{fig:dispersion_radii}, the stellar velocity dispersion-size plane which allows us to study the kinematic evolution of massive quiescent galaxies from $z=2-0$, is presented. The ongoing major merger galaxies are included to show that their incorrect composite dispersion measurement increase the scatter if not properly accounted for.

The mean velocity dispersion of the sample studied in this paper is $289\pm58$ km/s (without major mergers). This is consistent with previous $z>2$ massive quiescent galaxy literature (see studies shown in Figure \ref{fig:dispersion_radii}) with a mean dispersion of $272\pm31$ km/s. Our velocity dispersion and size measurements (including other structural parameters) for CP-1242752 (indicated by blue square in Figure \ref{fig:dispersion_radii}) are consistent with previously published values \citep{vandeSande2013,Belli+14,Kriek+16,Belli+17}.

Comparing the median dispersion of our study to that of the local MASSIVE(n) sample, a shallow or no kinematic evolution from $z=2-0$ is found. In Figure \ref{fig:dispersion_radii}, significant effective size evolution consistent with earlier findings are observed \citep{Newman+12,vanderWel+14}. The effect of the mass rank scatter on the fixed CND matching is shown as the purple shading around the median evolution. These shadings outline the variation on median when using upper and lower limit of the CND matching (based on the stellar mass cut ${\rm{log}}(M_{\ast}/M_{\odot})>11.70^{+0.07}_{-0.10}$) from the mass rank scatter.\\

Half of the morphologies of compact massive galaxies at $z\sim2$ have been suggested to be disk-dominated \citep{vanderWel+11}. So far only one spatially resolved study of a rotating disk quiescent galaxy at this epoch has been discovered \citep{Geier+13,Toft+17,Newman+18}. The line of sight measured velocity broadening of the absorption lines could be a combination of both rotation and dispersion in the presence of a disk-dominated system \citep[see an analytical prescription in][]{Belli+17}. Care must therefore be taken when comparing $z>2$ spatially unresolved dispersion with resolved local measurements.

\cite{Wuyts+11} shows that the stellar light distribution of galaxies, measured by the S\'ersic index, traces well the ${\rm{log}}(\rm{SFR})-{\rm{log}}(M_{\ast})$ relation, separating disk and spheroidal galaxies by $n=2.5$ at $z<1.5$. Under the assumption that this is valid at $z=2$, we classify our galaxies by S\'ersic index and find that $92\%$ of our galaxies have spheroidal ($n>2.5$) morphologies (when excluding the ongoing major mergers). If S\'ersic index $n>2.5$ is a good tracer of dispersion-dominated systems at $z>2$, it suggests that our sample of galaxy dispersion measurements are not strongly contaminated by rotation.

A recent study by \cite{Veale+18} presents the spatially resolved velocity dispersion measurements for the MASSIVE Survey sample. Here, ${\rm{log}}_{10}(M_{\ast}/M_{\odot})>11.7$ galaxies (similar to our stellar mass cut of the MASSIVE(n) sample) all have velocity dispersions in the range $200<\sigma<350$ km/s at all radii ($<15-30$ kpc). This rules out the possibility that the shallow dispersion evolution comparison is driven by spatial resolution. A comparison to the fixed CND-matched MASSIVE(n) sample establish that the dispersion remains nearly unchanged.

Negligible median dispersion evolution of our MQGs across the last $10$ billion years ($z=2-0$) is found in Figure \ref{fig:dispersion_radii}. In the absence of spatially resolved spectroscopy, we make use of the morphological classification which suggest that our kinematics are unlikely to be strongly contaminated by rotation. Studying the evolution of galaxies at fixed dispersion has been suggested as a method to minimise progenitor bias \citep[e.g.][]{Belli+14}.

\subsection{Stellar mass-size plane for massive quiescent galaxies} \label{sec:mass_size}

\begin{figure}
  \centering
    \includegraphics[width=8.5cm]{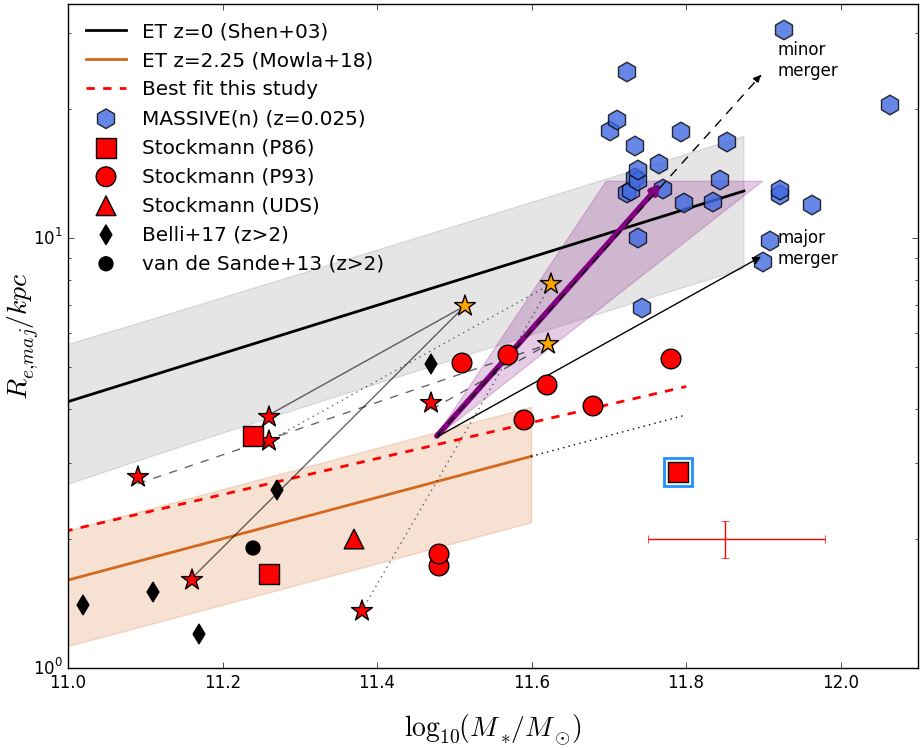}
  \caption{The stellar mass-size plane for massive, ${\rm{log}}_{10}(M_{\ast}/M_{\odot})>11.0$, quiescent galaxies: our sample (red symbols), other massive quiescent galaxies at $z>2$ \citep[black symbols,][]{vandeSande2013, Belli+17} and the MASSIVE(n) sample (blue hexagons). The representative error bar of our sample is shown in red. The source CP-1243752 is indicated with a blue square. The ongoing major merger-corrected stellar masses (red stars) are connected (gray dotted, dashed, and solid lines) to their post-merger positions (orange stars), following the \cite{Bezanson+09} prescription. The minor (dashed) and major (solid) merger-predicted evolutions from \cite{Bezanson+09} are shown with black arrows. The best fit relations at $z=0$ \citep{Shen+03} and $2.25$ \citep{Mowla+18}, with their $1\sigma$ uncertainty, are shown in black and brown, respectively. The best-fit relation to the galaxies of this study is shown in dashed red. The purple arrow shows the median evolution between our study and the MASSIVE(n) sample. The shaded purple area represents the uncertainty on the median of the MASSIVE(n) sample when the mass rank scatter from \cite{Behroozi+13} is taken into account (see explanation in Section \ref{sec:dispersion_size}). The median mass-size evolution of MQGs from $z=2-0$ can be explained primarily by minor mergers.}
  \label{fig:masssize}
\end{figure}

In Figure \ref{fig:masssize}, the stellar mass-size plane (${\rm{log}_{10}}M_{\ast} - R_{\rm{e,maj}}$) is presented which allows us to study the structural and stellar mass evolution of massive quiescent galaxies since $z\sim2$. The three ongoing major-merger galaxies with resolved sizes of the individual galaxies (Section \ref{sec:galfit_modeling}) and their flux corrected stellar masses (Section \ref{sec:major_merger_spectra}) are shown in the figure. The post-merger stellar masses and sizes of these are predicted using the argument of virialization from \cite{Bezanson+09}. The resulting position of post-merger galaxies is consistent with the average locus of the most massive (${\rm{log}}_{10}(M_{\ast}/M_{\odot})>11.5$) individual galaxies in our sample, showing that a way to form the most massive quiescent galaxies in our sample could be major quiescent-to-quiescent dry galaxy mergers \citep{Naab+06}.

A best fit relation to the galaxies in this study, including the major merger separated galaxies, reveal a shallower slope than what are found in \cite{vanderWel+14} $z=2.25$ mass-size relation, but in a better agreement with \cite{Mowla+18}. The best fit parameters, using a similar parametrization ($r/\rm{kpc}=A(M_{\ast}/(5\cdot10^{10}))^{\alpha}$), are ${\rm{log}}(A)=0.19$ and $\alpha=0.42$. The stellar mass for CP-1243752 (blue square in Figure \ref{fig:masssize}) is consistent within $1\sigma$ standard deviation with \cite{vandeSande2013} and \cite{Belli+17} and within $1.1\sigma$ for the stellar mass published in \cite{Kriek+16}.\\

The distribution of our sample shows that $z>2$ MQGs are $\sim2$ times more compact than objects with the same stellar mass in the local Universe \citep{Shen+03}, which is a well-established result in previous works \citep{vandeSande2013,Belli+17}. The median stellar mass and size for our (MASSIVE(n)) sample ${\rm{log}}(M_{\ast}/M_{\odot})=11.48\ (11.77)$ and $R_{\rm{e,maj}}/\rm{kpc}=3.42\ (13.55)$ show that a doubling ($\sim0.3$ dex) in stellar mass and a factor of $4$ in size evolution are required to bring the two samples into qualitative agreement.

Using the method from \cite{Bezanson+09} for predicting stellar mass and size growth, minor and major merger tracks are shown in the mass-size plane. The median mass-size evolution between our $z>2$ MQGs and the local MASSIVE(n) sample could be explained by minor merger-predicted size and stellar mass growth. The tracks start at the median size and stellar mass of our sample (only red symbols). The qualitative conclusions remain the same when using a mean instead of a median or changing the choice of reference (with/without the major merger galaxies).  

The median logarithmic mass-size slope is $\alpha=1.78^{+0.37}_{-0.29}$ ($r\propto{}M_{\ast}^{\alpha}$). The uncertainties are determined based on the CND mass rank scatter shown as the purple shaded area in Figure \ref{fig:masssize}. This confirms the suggestion that minor mergers ($\alpha=2$), compared to major mergers ($\alpha=1$), are the preferred evolutionary path in the mass-size plane. 

In line with earlier studies \citep{vandeSande2013,Belli+17,vanderWel+14,Mowla+18}, we find that our sample of $z>2$ MQGs is compact in the stellar mass-size plane and further suggests that minor merger-driven size evolution \citep{Bluck+12,Newman+12,Hilz+12,Hilz+13,Oogi+13,Fagioli+16} is preferred when comparing to the fixed CND-matched MASSIVE(n) sample.

\subsection{Stellar-dynamical mass plane for massive quiescent galaxies} \label{sec:dynamicalmass}

\begin{figure}
  \centering
  \includegraphics[width=8.5cm]{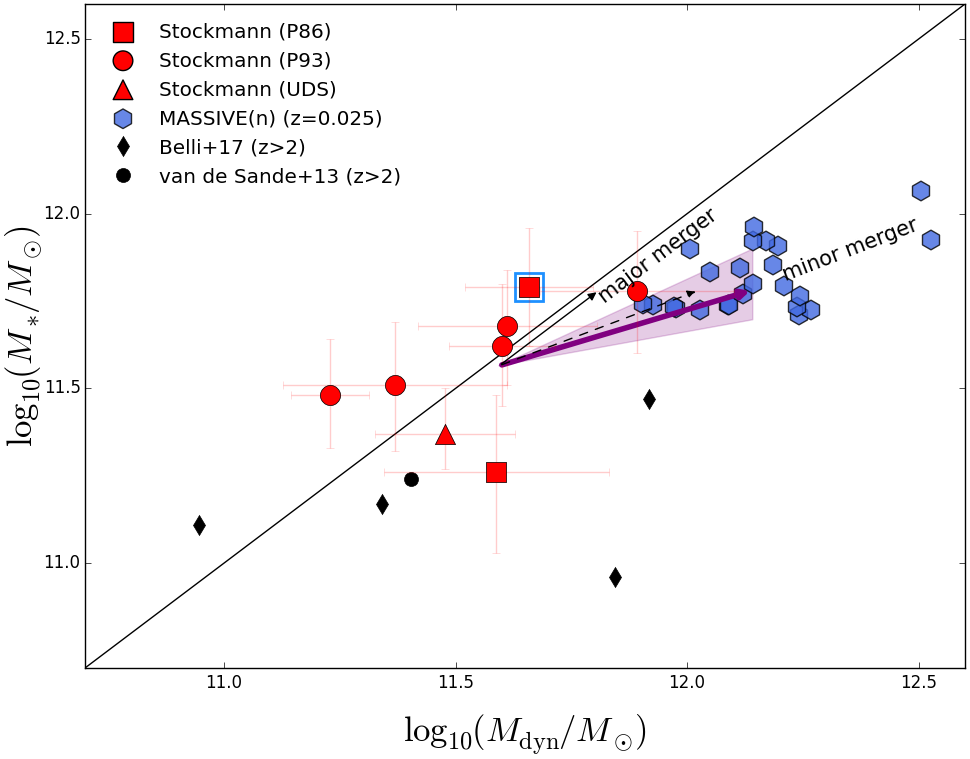}
     \caption{The dynamical-stellar mass plane for this study (red squares: \textit{P86}, circles: \textit{P93}, triangle: UDS19627), other $z_{\rm{spec}}>2$ massive quiescent galaxies \citep[in black symbols][]{vandeSande2013,Belli+17}, and the MASSIVE(n) sample (blue hexagons). The purple arrow connects the median of our sample with the median of the MASSIVE(n) sample. The purple shaded area represents the uncertainty on the median values of the MASSIVE(n) sample from the CND mass rank scatter (see explanation in Section \ref{sec:dispersion_size}). The solid black line is the $M_{\ast}=M_{\rm{dyn}}$ relation. The dashed/solid black arrow represents the predicted constant dispersion stellar-to-dynamical mass evolution for minor/major mergers \citep{Bezanson+09}. The blue square indicates the source CP-1243752 \citep[previously published in][]{vandeSande2013,Kriek+16,Belli+17}. The calculated dynamical-to-stellar mass ratio doubles from $z=2$ to $0$ when comparing to the fixed CND-matched MASSIVE(n) sample.}
  \label{fig:Mstar_Mdyn}
\end{figure}

In Figure \ref{fig:Mstar_Mdyn}, the dynamical-to-stellar mass relation for massive quiescent galaxies is plotted in order to study the interplay between the stellar and total (dynamical) mass potential over time. The dynamical mass derived from the Jeans equation \citep{Jeans+1902} for symmetrical systems is as follows:
\begin{equation} \label{eq:mdyn}
M(r) = \beta \frac{R_{\rm{e,maj}}\sigma^2}{G}.
\end{equation}
Here, $R_{\rm{e,maj}}$ is the effective semi-major axis, $\sigma$ is the stellar velocity dispersion, $G$ is the gravitational constant and $\beta$ is a parameter incorporating the full complexity of a collisionless systems with radial dependent parameters of density, dispersion, and velocity anisotropy. Following \cite{Cappellari_2006}, $\beta(n)= 8.87-0.831n+0.0241n^2$ is adopted where $n$ is the S\'ersic index \citep{Sersic}. The representation of $\beta$ is a good approximation for symmetric systems such as an elliptical galaxy that is well represented by a de Vaucouleurs profile. \cite{Taylor+10} and \cite{Cappellari+13} show that using such a parametrization of $\beta$ yields dynamical masses in better agreement with the stellar masses when the sizes, are estimated using a 2-dimensional S\'ersic fitting method, rather than a fixed value of $\beta$.

The galaxies of this study are consistent with the stellar-to-dynamical mass ratio, $M_{\ast}/M_{\rm{dyn}}<1$, within the large uncertainties. A ratio $>1$ is referred to as a non-physical (forbidden) region where the total mass is smaller than the mass of the stars. The galaxy, UV-230929, is located in this region at $1.1\sigma$ standard deviation from the $M_{\ast}/M_{\rm{dyn}}=1$ relation. Unfortunately, our large uncertainties prohibit trustworthy estimates of the total dust+gas mass for our sample. In \cite{Belli+17}, it is suggested that dispersion dominated systems with $n>2.5$ lie closer to the $M_{\ast}/M_{\rm{dyn}}=1$ relation at $z\sim2$. 

Compared to previous $z>2$ massive quiescent galaxy studies (see legend in Figure \ref{fig:Mstar_Mdyn}),  our sample occupies a similar dynamical mass range but has larger stellar masses. This is further discussed in Section \ref{sec:discussion_Mstar_Mdyn}. The dynamical mass for CP-1243752 (indicated by a blue square) is consistent with the previous measurements in \cite{vandeSande2013} and \cite{Belli+17}.\\

A comparison between our study with the MASSIVE(n) sample is made to learn about the fixed CND evolution in the dynamical-stellar mass plane. The median evolution in Figure \ref{fig:Mstar_Mdyn} illustrates that the dynamical mass evolves $2\times$ faster than stellar mass within the effective radii. This means that the galaxies evolve such that the $M_{\ast}/M_{\rm{dyn}}$ ratio decreases from $z=2$ to $0$. 

The minor and major merger evolution are shown for constant velocity dispersion evolution ($\Delta{r}\propto{}M_{\ast}^{\alpha}$), with $\alpha=1$ for major merger and $2$ for minor merger evolution. This is motivated by the shallow/constant dispersion evolution found in Section \ref{sec:dispersion_size}, when also comparing to the MASSIVE(n) sample. The median evolution from $z=2$ to present day prefers the minor merger predicted evolution when comparing our study to the MASSIVE(n) sample in the dynamical-stellar mass plane.

The median evolution from our study to the MASSIVE(n) sample at present day, in the dynamical-stellar mass plane, is consistent with minor merger evolution that is similar to what is found in Figure \ref{fig:dispersion_radii} and \ref{fig:masssize}.

\section{Discussion} \label{sec:sum_discussion}
The structural and kinematic evolution for massive galaxies from $z=2$ to present is explored by assuming that the galaxies, in this study, are the progenitors of the MASSIVE(n) sample. Such a claim has been motivated by a fixed CND-matching between the two samples of galaxies. This suggests that these galaxies undergo significant size growth together with shallow velocity dispersion evolution, driving up the dynamical-to-stellar mass ratio from $z=2$ to $0$. The role of major mergers in the evolution of massive galaxies is discussed following an interpretation using idealized and cosmological simulations. Furthermore, the origin of the dust heating, observed in the MIR and FIR emission, is discussed. Finally, the caveats are presented.

\subsection{Quiescent-to-quiescent major mergers}
Three galaxies in our sample, initially unresolved in ground-based imaging, are found in \textit{HST} images to be double sources and confirmed with X-shooter to be ongoing major merger systems (see Section \ref{sec:major_merger_spectra}). In this section, we discuss how the high major merger fraction (6/18) affects the prevalence of minor merger structural evolution of massive QG (found in Section \ref{sec:results}) and if the high fraction could be caused by a selection bias.

Following the definition in \cite{Man+12} we find a pair fraction of $20\pm12 \%$ when assuming no projected sources ($\langle{}N_{projected}\rangle=0$) and using the Poisson error estimate. In COSMOS and UDS a pair fraction of $10\ \%$ major mergers are found for massive (${\rm{log}}(M_{\ast}/M_{\odot})>11$) galaxies at $z=2$ \citep{Mundy+17}. In the case that the observed major mergers are representative for the complete sample of massive QGs we can estimate the number of major mergers each galaxy undergo ($N_{\rm{merger}}$) following the prescription in \cite{Man+16}. Under the assumption that the merger rate is constant from $z=2$ to $0$, equation (3) in \cite{Man+16} can be written as,  $N_{\rm{merger}}=\Delta{}t f_{\rm{pair}}/t_{{\rm{obs}}}$. The pair fraction, $f_{\rm{pair}}=0.2$, observation time $t_{{\rm{obs}}} = 0.8$ Gyr (from $z=1.9-2.5$) and time of evolution $\Delta{}t\sim10$ Gyr is used to estimate the number of mergers from $z=2$ to $0$. These numbers reproduce a major merger rate of $\sim1$ for a pair fraction of $10\ \%$ similar to what was suggested in \cite{Man+12}. For a $20\ \%$ pair fraction we find that on average each galaxy undergo $N_{\rm{merger}}=2.5$ mergers between $0<z<2$. This number of 1:1 major mergers would corresponds to a stellar mass increase of $0.5$ dex which is inconsistent with the stellar mass of the MASSIVE(n) sample (see Figure \ref{fig:masssize}). In the case of a $10\ \%$ pair fraction the stellar mass increase is consistent with the average stellar mass of the MASSIVE(n) sample, however in this case another mechanism must then be in place to produce the large size growth observed between our sample and the MASSIVE(n) galaxies. 

Our sample was selected to be UVJ quiescent and K-band bright, which could have introduced a bias for ground-based unresolved bright red systems like quiescent to quiescent galaxy major mergers (see also Section \ref{sec:caveats}). See also \cite{Mowla+18} and \cite{Marsan+18} that addresses the issue of close pairs of massive QG at $z\sim2$. If this selection bias is responsible for the high pair fractions, this could explain why we observe that the stellar mass-size evolution from $z=2$ to $0$ is dominated by minor mergers (see Figure \ref{fig:masssize}). The majority of our major merger targets are in the low stellar mass end of our sample (${\rm{log}}(M_{\ast}/M_{\odot})<11.5$). This could indicate that a possible way to produce ultra massive (${\rm{log}}(M_{\ast}/M_{\odot})>11.5$) QGs could be via quiescent-to-quiescent galaxy major mergers at $z>2$. A scenario involving early time major and late time minor merger evolution will be testable with larger samples of massive QGs at $z>2.5$.

\subsection{Minor-merger size evolution at constant dispersion} \label{sec:discussion_minormerger}

In Figure \ref{fig:masssize}, a slope of $\alpha=1.78^{+0.37}_{-0.29}$ is found for the mass-size evolution of our MQG from $z=2$ to $0$. Such an evolution can be interpreted using the analytical framework from \cite{Bezanson+09} and \cite{Naab+09} which find that minor merger-driven growth is needed to produce a mass-size slope of $\alpha=2$. An extended numerical treatment from \cite{Hilz+12} finds that when including the effect of escaping particles (a process arising from virialization following merger interaction), they recover a steeper mass-size slope ($\alpha=2.4$) alongside a constant dispersion evolution for minor merger-driven growth. Such a scenario could explain the observed size growth and shallow dispersion evolution observed.

The scenario presented in \cite{Hilz+12} occurs for two-component (stellar+halo) systems when they undergo 1:10 minor merger evolution. They reproduce the structural evolution found in \cite{Bezanson+09} and \cite{Naab+09} when simulating minor-merger evolution of stellar-only systems. According to \cite{Hilz+12} this suggests that the growth of the dark matter halo is an important ingredient necessary to cause the shallow dispersion evolution together with the expected size growth evolution we find in this study. Moreover, \cite{Hilz+12} shows that major mergers increase the dispersion and size proportional to the stellar mass. This is not what is found when comparing the size and dispersion evolution with the MASSIVE(n) sample (see Figure \ref{fig:dispersion_radii} and \ref{fig:masssize}). In the minor merger scenario, the velocity dispersion would be maintained in the inner region of the galaxy, as additional stellar mass is accreted in the outer parts from tidally stripped satellite systems. Over time, this would change the stellar light distribution on the outskirts of the galaxy, causing a continuous growth of the half-light radius \citep{vanDokkum+10,Hill+17}. 

In UV-105842, we may be observing a direct example of the minor merger-driven size increase. A small satellite system within close (spectroscopically confirmed) proximity of the central galaxy is found. Based on the flux-ratio estimated from the \textit{GALFIT} modeling we estimate a stellar mass ratio of 1:$12^{+6}_{-3}$ for this minor merger, consistent with the average 1:16 ratio estimated by \cite{Newman+12}. To double its stellar mass (as suggested by the median $\sim0.3$ dex increase derived for our sample), the galaxy would need to go through $\sim$ 12 such minor mergers between z=2 and 0. Other minor merger stellar mass ratios of 1:5, 1:10 and 1:20 suggested by \cite{Hilz+13} and \cite{Bedorf+13}, would correspond to $5$, $10$, and $20$ minor mergers between z=2 and 0 for a similar stellar mass increase. In \cite{Man+16} issues related to the translation of the H-band flux ratio to a stellar mass ratio (e.g. due to $M/L$ ratio variation in galaxies), directly affecting the above argument, are discussed.

Many observational \citep{Bluck+12,McLure+13,Fagioli+16,Matharu+19,Zahid+19} and numerical \citep{Naab+09,Oser+12,Oogi+13,Tapia+14,Naab+14,Remus+17} studies find that minor mergers could be a dominant process for the  size growth of massive galaxies, but it may not be able to explain the the full size evolution \citep{Cimatti+12, Newman+12}. Feedback processes have been shown to also affect the size growth \citep[e.g.][]{Lackner+12,Hirschmann+13}. Specifically AGN feedback is shown, by modern simulations, to be necessary to reproduce the observed size evolution \citep[see][]{Dubois+13,Choi+18}.


\subsection{Stellar-to-dynamical mass evolution} \label{sec:discussion_Mstar_Mdyn}

We found that the dynamical-to-stellar mass ratio shown in Figure \ref{fig:Mstar_Mdyn} increases by a factor of two within MQGs from $z=2$ to $0$. This could be attributed to either IMF changes of the stellar population \citep{Cappellari+12} affecting the stellar mass estimates or an increase in the dark matter fraction within the effective half-light radius. 

Numerical simulations find that minor merger-driven evolution alters the distribution of stars over time from a core to a core-envelope system by accretion of particles in the outskirts of the galaxy \citep{Hopkins+09,Hilz+12,Hilz+13,Frigo+17,Lagos+18}. A consequence of this is that the central dispersion remains constant while the half-light radius grows, encompassing a larger part of the dark matter halo and effectively increasing the dark matter fraction over time \citep{Hilz+12}.

A mass-size evolution similar to what we find is, according to \cite{Hilz+13}, caused by a massive dark matter halo that drives the accretion of dry (collisionless) minor mergers at large radii through tidal stripping. This inside-out growth increases the effective half-mass radius to encompass dark matter dominated regions which might explain the increase of the dynamical-to-stellar mass fraction within the half-light radius that we observe.

Care must be taken when interpreting the observations in terms of idealized numerical simulations. However, \cite{Remus+17} also find that the central dark matter fraction increases with decreasing redshift when comparing different cosmological simulations. Furthermore, observational evidence for inside-out growth in massive galaxies is presented in \cite{Szomoru+12}.

In Figure \ref{fig:Mstar_Mdyn}, we find that our sample is consistent with the dynamical-to-stellar mass ratio of one suggesting low dark matter fractions at $z\sim2$. For a stellar mass increase of $0.3$ dex (similar to our median evolution), \cite{Hilz+12} predict a dark matter fraction increase of $\sim70\ \%$ within the effective radius. If we assume that the mass of the galaxy consists only of dark matter and stars, we can estimate the dark matter mass fractions ($M_{DM}/M_{\rm{dyn}} = 1 - M_{\ast}/M_{\rm{dyn}}$), from the dynamical-to-stellar median ratio at $z=2$ and $0$, to be $7_{-7}^{+24}\ \%$ and $56\pm8\ \%$, respectively. This suggests an increase of the dark matter fraction within the effective radius of $17-64\ \%$. Note, however, that this increase cannot purely be associated with the dark matter from the minor mergers as the growing half-light radius similarly encompasses more of the central dark matter halo and also contributes to this increase.

According to \cite{Remus+17}, the mass growth of massive galaxies can be explained by two stages: 1) High redshift in situ mass growth resulting in a dense stellar component in the center of the potential where the dark matter fraction is low, 2) dry merger events dominate the mass growth at lower redshift (with major mergers being rare) resulting in the build-up of a stellar envelope increasing the half-light radius and thus the dark matter fraction (similar to the interpretation above).


\subsection{Dust heating in massive quiescent galaxies at $z>2$} \label{sec:disc_agn}
The $24\ \mu{m}$ SFR limit, used to restrict the stellar population models, results in specific SFRs for our galaxies of ${\rm{log}}_{10}(\rm{sSFR/yr})<-10$. Nonetheless, stronger limits on the specific SFR can be obtained if the source of dust heating is not caused by recent star formation. In Section \ref{sec:SFR_tracers}, the information from optical nebular emission and mid-IR is combined to set stringent limits on the SFR of our sample (see also Figure \ref{fig:SFR_Mstar}). This information reveals that our sample lies $1.5$ dex below $z=2$ the star formation - stellar mass relation of \citep{Speagle+14} (extrapolated to  ${\rm{log}}_{10}(M_{\odot}/M_{\ast})\sim11.5$). 

Low-luminosity AGN is shown to be common in massive, ${\rm{log}}_{10}(M_{\ast}/M_{\odot})>11$, quiescent galaxies at $z<1.5$, through excess radio emission in stacked samples \citep{Man+16a, Gobat+18}. Six galaxies, in our sample, have direct radio detections; three of them with matching mid-IR detections (see Table \ref{tab:specz_SEDparam}). This could be evidence in line with the results from \cite{Olsen+13} who find a high fraction of AGN in massive quiescent galaxies at $1.5<z<2.5$ using X-ray stacking. Low luminosity AGN activity has, in \cite{Schawinski+09,Best_Heckman+12}, been associated with the suppression of SF which is an important effect in maintaining galaxies quiescent. Low levels of dust heating have also been associated with evolved stellar populations as a significant source to emit at wavelengths beyond $>160\ \mu{m}$ \citep{Salim+09,Bendo+12,Fumagalli+14,Utomo+14}. However, with no detections in the Herschel/PACS bands, we cannot rule this scenario out. In the case where AGN are indeed the dominant dust heating source in the galaxies, we can expect that the $24\ \mu{m}$ flux does not arise from residual SF. This is consistent with \cite{Whitaker+17} that find no strongly obscured SF in massive quiescent galaxies at $z>2$. Assuming the $24\ \mu{m}$ emission is not due to obscured starformation,  we find a specific SFR, ${\rm{log}}_{10}(\rm{sSFR/yr})<-11$, based purely on the optical emission limits/detections. The MIR-to-radio emission of the sample will, in a future publication, be investigated in detail (Cortzen et al. in prep).


\subsection{Caveats} \label{sec:caveats}
The main limitations of the results are here presented in bullet points:

\begin{itemize}
    \setlength\itemsep{0em}
    \item Overestimated stellar masses would lead to a shallower mass-size evolution and dynamical-to-stellar mass ratio evolution. Nonetheless, substantially overestimated stellar masses are ruled out by our dynamical masses being in agreement with previous kinematic studies of massive quiescent galaxies at $z>2$ \citep{Toft+12,vandeSande2013,Bezanson+13,Belli+14,Belli+17}.
    
    \item If rotation is significant in massive quiescent galaxies at $z>2$, the measured velocity dispersion, depending on the inclination, could have an unknown contribution from rotation resulting in heightened dispersion measurements. On the other hand, dispersion measurements from face-on rotation-dominated galaxies could result in low values. This would further drive the dynamical mass artificially down. Such issues should be addressed by spatially resolved spectroscopy where the $V_{rot}/\sigma$ can be estimated. 
    
    \item Previous studies \citep{Mancini+10} have suggested that sizes might be underestimated due to non-detection of low luminosity profile wings. However, ultra-deep imaging out to many effective radii does not find that this is the case \citep{Szomoru+10,Szomoru+11}.

    \item Dynamical-to-stellar mass evolution is sensitive to the determination of $\beta(n)$. The prescription from \cite{Cappellari_2006} is used, yet, this relation is determined from local galaxies and is assumed to be representative for dynamical systems at $z\sim2$. When comparing with the MASSIVE(n) sample, we assume a S\'ersic index of $n=4$, to be a fair representation of a spheroidal system. When changing the choice of $\beta=2-6$ for the MASSIVE(n) sample, the conclusion that the ratio must evolve from $z=2-0$ remains.
    
    \item The sample is $60\ \%$ mass complete for the massive (${\rm{log}}_{10}(M_{\ast}/M_{\odot})>11$) and K-band brightest ($K<20.5$) UVJ quiescent galaxies at $1.9<z<2.5$. This selection depends strongly on the performance of the photometric redshift estimate. In Figure \ref{fig:redshift}, we show that this works well for our sample using the catalog from \cite{Muzzin_COSMOS_Uvista}. This suggests that the sample studied in this paper is representative of the selection we presented in Section \ref{sec:sample_selection}. However, the photometry is used to select red systems and, consequently, introduce a selection bias towards mergers between red galaxies. An unresolved merger of a quiescent galaxy with a star-forming galaxy would produce a resulting bluer system that might be excluded from the selection.

\end{itemize}


\section{Summary and Conclusion} \label{sec:summary}

We examined the largest sample of massive quiescent galaxies observed to date at $z>2$ with deep X-shooter spectroscopy and HST/WFC3 imaging. We extend previous searches for very massive quiescent galaxies at $z>2$ to the K-band brightest \textit{UVJ} quiescent galaxies in COSMOS \citep{Muzzin_COSMOS_Uvista}, constructing a sample of $15$ MQGs. Full SED modeling of the photometry and spectroscopy confirms the sample to be $\sim1.5$ Gyr old, massive, ${\rm{log}}_{10}(M_{\ast}/M_{\odot})>11$, quiescent galaxies. 3 out of 15 galaxies are confirmed as ongoing major merger using both imaging and spectroscopy. In total, $40\ \%$ of the sample show evidence of mergers (minor or major) or other disturbed morphologies in HST/WFC3 $H_{F160W}$ imaging, suggestive of ongoing morphological transformation. The morphological information is used to correct the stellar masses prior to comparing the stellar populations, kinematics and structure/morphology of the galaxies to the MASSIVE(n) sample. We list below the main conclusions of the paper:\\

\begin{itemize}
    \setlength\itemsep{0em}
    \item We find that our galaxies lie $1-1.5$ dex below the extrapolation at the high stellar mass end of the SFR main-sequence \citep{Speagle+14} at $z=2$ and can be considered quiescent with low specific SFR, ${\rm{log}}_{10}(\rm{sSFR/yr})<-10.5$. These limits are based on optical emission line and MIR emission limits and detections. $1/3$ of the galaxies are detected in the MIR which could be caused by residual SF. However, more than half of our sample ($60\ \%$ of the MIR detections) have radio emission detected at $1.4$ or $3$ GHz. This radio emission is likely associated with AGN activity, a proposed heating mechanism leading to quenching and/or the maintenance of quiescence in massive galaxies.

    \item We find indirect evidence pointing to our velocity dispersion measurements to be minimally contaminated by rotation. Our systems also have a S\'ersic index $n>2.5$ (see Section \ref{sec:dispersion_size}). A direct comparison between our study and the MASSIVE(n) sample, shows evidence for shallow or no velocity dispersion evolution from $z=2-0$.
    
    \item Our sample is compact, in line with previous studies at $z\sim2$ \citep{vanderWel+14,Mowla+18}. We find that the median mass-size evolution ($\Delta{}r\propto{}\Delta{}M_{\ast}^{\alpha}$) compared to the MASSIVE(n) sample is best described by $\alpha=1.78^{+0.37}_{-0.29}$. This is consistent with both the simple kinematic predictions of minor merger driven size evolution from \cite{Bezanson+09} and the more extensive numerical treatment from \cite{Hilz+12}.
    
    \item We find that our sample of $z>2$ MQGs is consistent with a dynamical-to-stellar mass ratio $M_{\ast}/M_{\rm{dyn}}<1$ but that the shallow dispersion and significant size increase lead to an increasing dynamical-to-stellar mass ratio, doubling from $z=2$ to the present day. Such an effect is shown to be reproduced for an increasing dark matter fraction from $z=2-0$, within the effective radius of the galaxy \citep{Hilz+12}.
\end{itemize}

In this paper the largest sample of MQGs at $z>2$ with kinematic and structural observations, found via the mass-size and dynamical-stellar mass plane, is presented. A fixed CND-matching suggests that our sample of galaxies are the progenitors of the most massive and oldest elliptical galaxies in the local Universe, thus connecting $10$ billion years of evolution. These galaxies show a broad range of disturbed morphologies, confirming that mergers play a significant role in their morphological transformation and evolution to $z=0$.

In a companion paper, the relationship between the size and dispersion will be explored by studying the Fundamental Plane at $z\sim2$ and its consequent evolution to the present-day Universe (Stockmann+19b in prep).\\

We thank the anonymous referee for a constructive report that helped us improve the quality of the manuscript. We thank Martin Sparre for his useful discussions related to X-shooter data. M.S. extend gratitude to Nina Voit for her ultimate support and patience in the becoming of this work. Based on data products from observations made with ESO Telescopes at the La Silla Paranal Observatories under ESO programmes ID 086.B-0955(A) and 093.B-0627(A) and on data products produced by TERAPIX and the Cambridge Astronomy survey Unit on behalf of the UltraVISTA consortium. M.S., S.T., G.M., C.G., G.B., and C.S. acknowledge support from the European Research Council (ERC) Consolidator Grant funding scheme (project ConTExt, grant number 648179). The Cosmic Dawn Center (DAWN) is funded by the Danish National Research Foundation under grant No. 140. Based on observations made with the NASA/ESA Hubble Space Telescope, obtained from the data archive at the Space Telescope Science Institute. STScI is operated by the Association of Universities for Research in Astronomy, Inc. under NASA contract NAS 5-26555. Support for this work was provided by NASA through grant number HST-GO-14721.002 from the Space Telescope Science Institute, which is operated by AURA, Inc., under NASA contract NAS 5-26555. This research made use of Astropy (version 1.1.1),\footnote{http://www.astropy.org} a community-developed core Python package for Astronomy \citep{astropy:2013, astropy:2018}. This research made use of APLpy, an open-source plotting package for Python \citep{aplpy+12}. I.J. is supported by the Gemini Observatory, which is operated by the Association of Universities for Research in Astronomy, Inc., on behalf of the international Gemini partnership of Argentina, Brazil, Canada, Chile, the Republic of Korea, and the United States of America. GEM acknowledges support from  the Villum Fonden research grant 13160 “Gas to stars, stars to dust: tracing star formation across cosmic time”, the Cosmic Dawn Center is funded by the Danish National Research Foundation. A.M. is supported by the Dunlap Fellowship through an endowment established by the David Dunlap family and the University of Toronto. R.D. gratefully acknowledges support from the Chilean Centro
de Excelencia en Astrof\'isica y Tecnolog\'ias Afines (CATA) BASAL grant AFB-170002. M. H. acknowledges financial support from the Carlsberg Foundation via a Semper Ardens grant (CF15-0384). Y.P. acknowledges NSFC Grant No. 11773001 and National Key R$\&$D Program of China Grant 2016YFA0400702.


\appendix


\section{Further details on the reduction of the images} 
\subsection{PSF \& astrometry} \label{sec:hst_reduction_matching}
The $H_{F160W}$ images from our program and the ancillary COSMOS $F814W$ images employed in this work do not share the same World Coordinate System (WCS). We need to guarantee that the astrometry is common and accurate in both bands. Therefore, we chose to align the images to the COSMOS ACS $F814W$ image as the reference frame, which is registered to the fundamental astrometric frame of the COSMOS field, ensuring an absolute astrometric accuracy of 0\farcs05--0\farcs1 or better. Following \cite{Gomez_Guijarro+18}, we use \texttt{TweakReg} along with SExtractor \citep{SExtractor} catalogs of the two bands with the $F814W$ catalog and frame as references to register the images. After this, the images in both bands are resampled to a common grid and a pixel scale of 0\farcs06\,pix$^{-1}$ using SWarp \citep{2002ASPC..281..228B}. In addition, the spatial resolution of the two \textit{HST} bands is also different. Following \cite{Gomez_Guijarro+18}, we degrade the $F814W$ to the resolution of the $F160W$ data (0\farcs18 FWHM). We calculate the kernel to match the ACS $F814W$ to the PSF in the $F160W$ images employing the task \texttt{PSFMATCH} in IRAF, including a cosine bell function tapered in frequency space to avoid introducing artifacts in the resulting kernel from the highest frequencies. Then, we convolve this kernel to the $F814W$ image to achieve a common spatial resolution.

\subsection{Modeling of foreground and background sources} \label{sec:foreground_background}
Based on the spatially offset emission in the 2-D X-shooter spectra, we determine if candidate sources are within close proximity to the central galaxy. In Figure \ref{fig:HST_presentation} the central sources along their spatially offset sources is shown. UV-171687 shows offset H$\alpha$ and [NII] emission arising from a south-western source that we establish to be a foreground galaxy at $z=1.51$. We find another foreground galaxy north-east of UV-171060 at $z=1.37$ based on assuming that the single emission line detection is H$\alpha$. North-east of UV-155853 we find a background galaxy at $z=2.36$ (best visible in the Galfit modeling residuals of Figure \ref{fig:HST_presentation}) determined from the [OIII] doublet at $4959,5007$ \AA{}. For UV-105842 we find two spatially offset source, 1) $\sim3''$ north-east and 2) $\sim1''$ north-east. Source 1) is a foreground galaxy at $z=0.44$ based on the detected strong O[III] doublet at $4959,5007$ \AA{} and H$\alpha$ emission. For source 2) we find the [OII] doublet at $3726.2,3728.9$ \AA{}, O[III] doublet at $4959,5007$ \AA{}, and H$\alpha$ corresponding to a redshift $z=2.0124$. The latter redshift corresponds to a velocity offset of $2130\pm120$ km/s (uncertainty is calculated based on the spread of the individual redshift measurements) suggesting that it is not gravitationally bound to the central galaxy. Another option could be an offset AGN with high peculiar velocity following a merger.


\section{Details on modeling of the velocity dispersion} \label{app:dispersion_test}

\subsection{Statistical and systematic uncertainties}
To estimate the statistical error we measure the spread of the velocity dispersion distribution obtained from running pPXF on a 1000 data realizations. The data realizations are made by perturbing the pPXF best fit model with the pipeline estimated error spectrum, by linearly drawing values from a Gaussian with a mean of zero and spread of the initial errors. The X-shooter pipeline-estimated noise map is subjected to a wavelength dependent correlation of the pixels. We take this effect into account by scaling our noise spectrum to a reduced $\chi_{red}^2 = 1$ (assuming the errors are Gaussian). We follow the method used in \cite{Toft+17} and fit a 2nd order polynomial to a 50 pixel running reduced $\chi^2$ that we use to make a correction noise map, $\sigma_{\chi^2_{corr}} = \sigma_{\chi^2_{original}} \sqrt{\chi^2_{fit}}$.\\
We estimate the systematic error by testing how the dispersion is changing with the correction polynomial and implemented wavelength range.
We construct a grid of correction polynomials up to 24th order of both additive and multiplicative polynomials, where we find an average of $20\ \%$ variation from the fiducial dispersion, except for UV-239220, UV-773654, UV-171060, and CP-1291751.
When varying the start wavelength range ($[\lambda_{start},\lambda_{end}]$) within the interval $[3750-4050, 5950]$ and the end wavelength within the interval $[3750,4050-5950]$, we find that overall the dispersions are stable. In a few cases, the velocity dispersion increases well above the median dispersion (with varying wavelengths) with $50-100\ \%$ when excluding the higher order Balmer and Ca H+K lines, highlighting their importance. When including the end wavelength $\lambda>4500$ we find more stable dispersion measurements, not surprising as otherwise only half of the spectrum is included. The low S/N cases have more unstable dispersion values when excluding wavelength areas, highlighting the importance of understanding the systematic uncertainties. We sum up the wavelength and polynomial test by confirming that our fiducial velocity dispersions are robust (except for UV-239220, UV-773654, UV-171060, and CP-1291751). The systematic error is primarily due to template mismatch and as a result, we estimate the systematic error from the minimum and maximum values of the dispersion when using the full wavelength range and varying the additive and multiplicative correction polynomials, $\sigma_{sys} = 2/3\cdot(\sigma_{max}-\sigma_{min})/2$. This method is subjected to catastrophic outliers, and prior to the systematic error estimate, we exclude dispersion values more than $5\sigma$ outside of a Gaussian mean. We find that the systematic errors are on the order of the statistical uncertainties.\\

\subsubsection{Additional tests}
We measure the dispersion while excluding a window of $1600$ km/s along the wavelength direction in steps of $5$ \AA{}, to test whether the measured dispersion is dominated by specific lines. We find that the fiducial dispersion is very stable against excluding individual lines, and did not find a consistent decrease in the velocity dispersion similar to previous studies when excluding the H$\beta$ line \citep{vandeSande2013,Toft+17}. We allow pPXF to construct a linear combination of templates from the stellar library of BC03 with a Chabrier IMF and solar metallicity and find similar redshifts and velocity dispersions as our fiducial values which are reassuring.


\section{Details on the emission line fitting} \label{app:on_source_emission}

For UV-108899, we find that when fitting a double Gaussian profile to [OII] ($3726+3729$ \AA{}) fixed to the redshift of the central galaxy, gives the most conservative (highest) flux estimate. We try fitting with a single profile while using a free redshift parameter but recover high $\chi^{2}$ solutions. We list this conservative flux estimate, corresponding to a $\rm{SFR}=6\pm4\ \rm{M_{\odot}/yr}$ \citep{Kennicutt1998}, in Table. \ref{tab:specz_SEDparam}.

For UV-239220, we detect excess emission in the region of H$\alpha$ and the [NII] ($6548+6583$ \AA{}) doublet. With a fixed ratio between the [NII] doublet, we try three types of triple Gaussian profile models (free redshift+dispersion limit of $250$ km/s, free redshift+dispersion limit of $1000$ km/s, and fixed redshift+dispersion limit of $1000$ km/s) that all result in $\chi^2>2.4$ with no preferred solution. If we assign all of the flux in the excess to H$\alpha$ we obtain a conservative \cite{Kennicutt1998} SFR upper limit of $\sim30\ M_{\odot}/\rm{yr}$ (${\rm{log}}(sSFR)<-10\ \rm{[yr^{-1}]}$) consistent with the FIR and rest-frame optical upper limits from Section \ref{sec:SFR_uplim}. This confirms that the galaxy has low specific star formation consistent with its selection.


\begin{figure*}
  \centering
    \includegraphics[width=18cm]{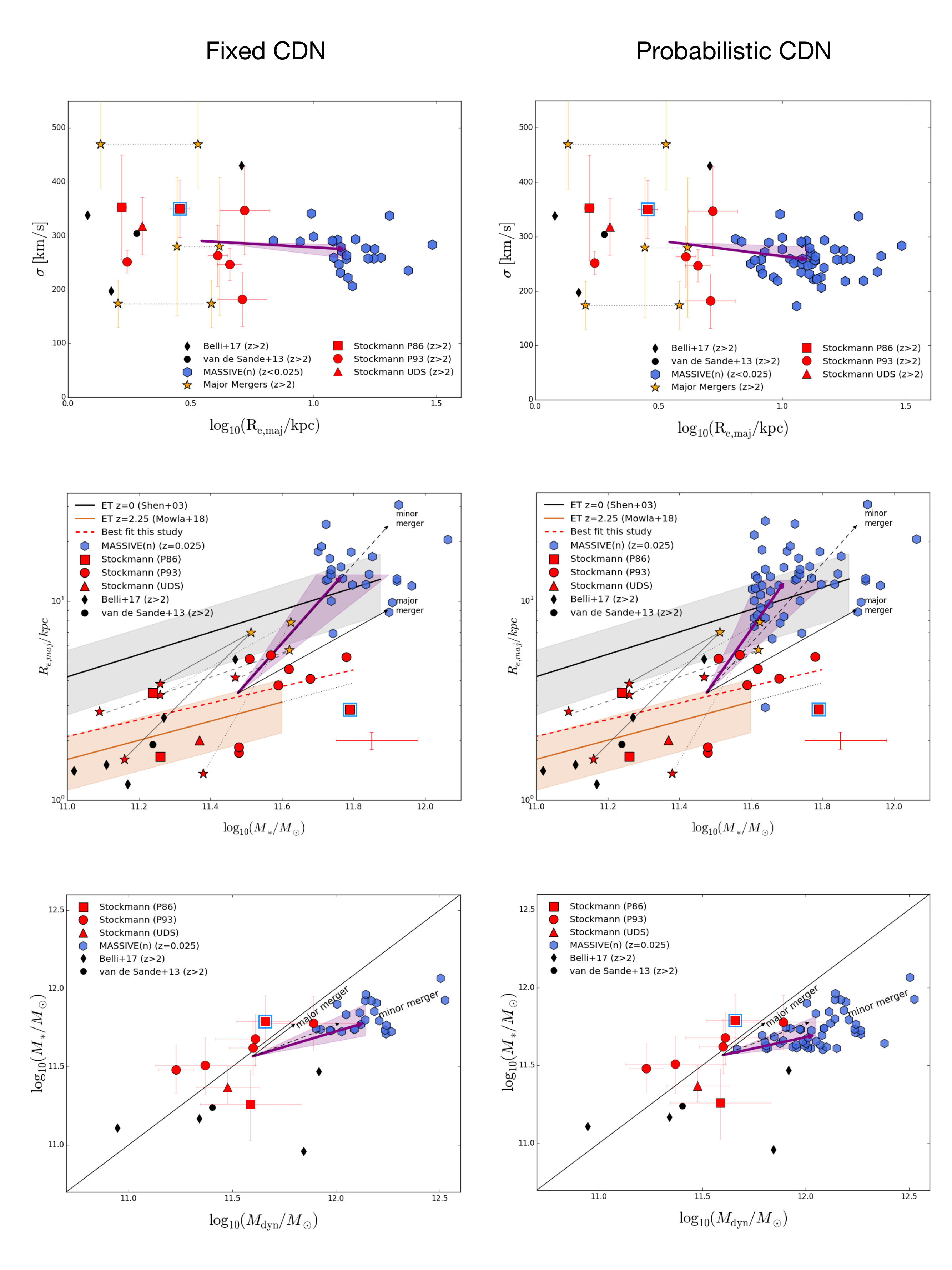}
  \caption{The figures from Section \ref{sec:results} are shown with the fixed and probabilistic \citep{Wellons&Torrey+17} CND-match to the MASSIVE Survey. For each of these methods our qualitative conclusions remain.}
  \label{fig:result_comparison}
\end{figure*}

\section{Comparing different CND methods} \label{app:CND}
The MASSIVE(n) sample was established using the assumption of a fixed CND from $z=2$ to $0$. To show that our results are robust against the choice of CND matching method we show the three result figures from Section \ref{sec:results} in Figure \ref{fig:result_comparison} using both the fixed CND matching and the probabilistic approach presented in \cite{Wellons&Torrey+17}.\\

The probabilistic approach uses numerical simulations (e.g. Illustris) to estimate the probability that a galaxy at $z=0$ are the descendant of a galaxy at redshift, $z_{obs}$. This method therefore allows to predict the most probable CND at $z=0$ for a population of galaxies with the specific CND at $z=2$ following the evolution of a numerical simulation. This method is thus a different approach than the fixed CND approach and in Figure \ref{fig:result_comparison} we show that adopting these two methods of connecting galaxies across time leads to the same conclusions.


\bibliographystyle{aasjournal}{}
\bibliography{PaperI_MQGz2}{}


\end{document}